\documentclass[aps,pre,preprint,groupedaddress]{revtex4}
\usepackage{graphicx}
\usepackage{amsmath}
\usepackage{bm}
\usepackage{epsf,epsfig,graphics}
\usepackage{float}
\usepackage{makeidx}
\usepackage{latexsym}
\usepackage{amssymb}
\usepackage{mathrsfs}
\usepackage{amsbsy}
\usepackage{subfigure}
\usepackage{CJKutf8}

\newcommand{\figws}{7.5cm}
\newcommand{\figwss}{5.cm}

\usepackage{color}
\usepackage{morefloats}
\DeclareMathAlphabet\mathbfcal{OMS}{cmsy}{b}{n}

\begin{document}
\draft
%\twocolumn[\hsize\textwidth\columnwidth\hsize\csname@twocolumnfalse\endcsname

\title{Optimized two-dimensional Networks with edge crossing cost: frustrated anti-ferromagnetic spin system}
\author{An-Liang Cheng\begin{CJK*}{UTF8}{bsmi}(鄭安良)\end{CJK*} and Pik-Yin Lai\footnote{Email: pylai@phy.ncu.edu.tw}\begin{CJK*}{UTF8}{bsmi}(黎璧賢)\end{CJK*} } 
\affiliation{Dept. of Physics and Center
for Complex Systems, National Central University, Chung-Li District, Taoyuan City 320, Taiwan, R.O.C.}

\hfill\today

\begin{abstract}
 We  consider a quasi two-dimensional
network connection growth model that minimizes the wiring
cost while maximizing the network connections, but at the same time edge crossings are penalized or forbidden. This model is mapped to a dilute anti-ferromagnetic Ising spin system with frustrations. We
obtain analytic results for the order-parameter or mean degree of the optimized network using  mean field
theories. The cost landscape is analyzed in detail showing complex structures due to frustration as the  crossing penalty increases. For the case of strictly no edge crossing is allowed, the mean-field equations lead to a new algorithm that can effectively find the (near) optimal solution even for this strongly frustrated system. All these results are also verified by Monte Carlo
simulations and numerical solution of the mean-field equations. Possible applications and relation to the planar triangulation problem is also discussed.
\end{abstract}
%\pacs{ }
%]
\maketitle
\section{Introduction}
 Due to the availability of a large number of scientific, technological, social, and financial data, research related to complex networks have been vastly expanding  in the last couple of decades\cite{bigdata,newmanbook}. And the convenience the data banks accessible through internet further encouraged the intensity of research in these areas\cite{newman,strogatz}. Several issues are of special interest, one is how the network topology affects the function of these complex networks, as in the case of biological networks.
% Since mere information about the static structure of networks cannot reveal the entire relations or interactions among the nodes and the dynamical data of network nodes are increasingly available, there are efforts concerning the reconstruction of  networks from the time-series dynamics data of the nodes\cite{CLL2013,CLL2015,zhang2015,CT2017,Lai2017,TCL,Lai2019}.
 In order to grasp the fundamental understanding of the structure and functions of complex networks,  the principles behind for the formation or growth of such networks need to be investigated. This question is associated with evolution of the network which involves the growth of nodes and connections.  Presumably the system is optimizing towards some goal\cite{chaosfocus,MWong2017}, but often  constrained by some external or internal conditions and also regulated by the feedback of the status of the network. This is related to the selection and adaptation in biology in a broad sense\cite{Katifori2016,jyhuang,sporn}.
 Also  there are some recent efforts on attempting to understand the designing principle for network that optimize information flow\cite{bialek2009,bialek2010} or robustness against attacks\cite{Havlin2005,Bornholdt2012}.

Recently,  we developed a theoretical framework for such optimized  networks and investigate the connection formation models of some physical complex networks or bio-networks. Our motivation is to understand the relation between the energy cost and the connectivity in a network. We consider undirected network connection growing models that aim at minimizing the cost of connections while at the same time maximizing the network connections or other important network properties. Classical network models and statistical mechanics methods were employed to study the growth of connections in networks under the principle of optimizing some objective function\cite{cheng}. These models are quite general and applications could be found in various physical or biological networks. By mapping the system to a spin model, it was demonstrated that some version of these models can be solved exactly, and in several cases  exhibiting interesting phase transition behaviours\cite{cheng}.  On the other hand, the growth of connections in a network is often  subjected to some possible constraints that  strongly affect the resultant structure of the optimized network. This is especially relevant for physical network connections in which the connecting links exhibit mutual interactions. Such constraints or interactions can lead to frustrations in the course of finding the optimized network which is a challenging problem, both theoretically and computationally.

In this paper, we consider nodes embedded in a two-dimensional plane and the network connections are subjected to interactions of intersecting links or even under the constraint of no edge crossing. This optimized  network connection problem has a broad applications on physical networks such as the designing principle in a two-dimensional electric network, which is still an active research area.  A circuit network with minimal wire cost but can connect a large number of nodes would be desirable.  However, no bond crossing is a necessary condition for
electric circuit board network since crossing of bare metallic wires will lead to a short circuit. The topological constraint\cite{doi,sheng98} of no crossing between two strands is rather general and gives rise to the entanglement interactions in polymer melts\cite{doi} and knots\cite{huang01,Lai02}.
The constraint of no bond-crossing will have a strong effect on the structure of the resulting two-dimensional networks.  One can further generalized to quasi two-dimensional networks in which edge crossings are penalized or forbidden; representing an extra cost of insulation at the wire crossing junctionfor the case of an electric circuit. This model is also of relevance to the growth of cultured neuronal network in vitro\cite{JiaPRL2004,LaiJia2006} or blood vessels in tissues.
As will be shown in this work, the edge crossing penalty corresponds to the  frustrated anti-ferromagnetic spin system and for the strongly edge crossing interacting case,  there are many  near local optimal solutions  in the low-temperature regime that hinder the finding the true optimized network.
 In this paper, we derive analytic results using  the mean field theories and also verified these results explicitly by  Monte Carlo simulations and numerical solutions of the mean-field equations. For the challenging case of no bond-crossing characterized by strong frustrations, we further develop an efficient algorithm to find the optimal network configurations.

\section{Edge crossing penalty in 2D networks: dilute anti-ferromagnetic Ising model and the mean-field equations}
Consider an undirected network of $N$ nodes which are embedded in a two-dimensional plane, the connections are
described by the (symmetric) adjacency matrix
 $A_{ij}$  which takes the value of 1 if there is an edge
connecting between nodes $i$ and $j$ and 0 otherwise, and there is no self-connection ($A_{ii}=0$).  In
general, each edge $A_{ij}$ is associated with a  weighing factor
denoted by $w_{ij}$ which can be interpreted as the energy or
cost required to set up such a connection.
The maximal total  number of possible edges is
$N_b\equiv N(N-1)/2$.
 For convenience, the edge of the network is labeled by the
Greek index $\alpha$ ($\alpha=1,2,\cdots, N_b)$ hereafter.
The average over the entire network  is denoted by an overbar, so the mean degree is ${\bar k}\equiv{1\over N}\sum_{i=1}^N k_i$. The average can also be over the edges and in this case $\overline{\cdots}\equiv {1\over{N_b}}\sum_{\alpha=1}^{N_b}\cdots$.
 By introducing
the Ising spin variables
\begin{equation} S_{\alpha}\equiv 2 A_\alpha -1,
\end{equation}
which takes values $\pm 1$, the links of the network can be described in terms of the Ising spins. The order-parameter (magnetization in the spin system), denoted by $x_o$, is
\begin{equation}
x_o=\frac{1}{N_b}\langle\sum_{\alpha=1}^{N_b}S_{\alpha}\rangle,
\end{equation}
where the thermodynamic (ensemble) average is denoted by $\langle \cdots \rangle$.
The  mean degree of the network, ${\bar k}$, is related to $\{S_\alpha\}$ by
${\bar k}={{N-1}\over 2}+\frac{1}{N}\sum_{\alpha}S_{\alpha}$.  $x_o=1$
and -1 corresponds to the case of completely connected and
unconnected networks respectively. The ensemble average of the
  mean network degree is related to the
 order-parameter and average mean connectivity via
\begin{equation} \frac{\langle{\bar k}\rangle}{N-1}=\langle{\bar A}\rangle=\frac{1+x_o}{2}.\label{meank}\end{equation}
 Thus by introducing a temperature (which corresponds to the noise level of the system), the problem of finding the fully optimized network is mapped to the search of the ground state of the corresponding Ising spin system.
In general, the spin system can be conveniently simulated by Monte Carlo method and annealed down to low temperatures to obtain the (near) ground state optimized network configuration.

Here we consider a network embedded in a two-dimensional plane with
physical connections. The cost of constructing the network would
include the material cost of the edge,  which is simply proportional to the distances
between two nodes (denoted by the matrix $d_{ij}$), and the cost of making the connections which can obey some probability distribution but is taken to be a fixed constant $c_o$ (independent of the separation of the nodes)  in this paper for simplicity. The connection cost between  nodes $i$ and $j$ is denoted by the weight $w_{ij}=d_{ij}+c_o$ and enters in the first term in Eq. (\ref{HC}) below.

 Similar to Model A in \cite{cheng}, one wishes to design a network that minimizes the connection cost and at the same time maximizes the network connections. However
the in-plane confinement of the nodes would unavoidably lead to the 
consideration of possible edge crossings, and in general edge
crossings would increase the total cost of the network. Examples include the
insulation for metal wires for crossings in electrical circuit chips and the construction of flyover or traffic junctions when road crossing occurs in road traffic.
 To model the
effect of the bond crossing, an edge-crossing cost term (last term in (\ref{HC})  is included in the total cost (${\cal C}_X$) of the model as
\begin{equation}
{\cal C}_X = \sum_{\alpha}A_{\alpha}w_{\alpha} -\lambda N{k}+\frac{4 \gamma}{N_b} \sum_{\alpha'<\alpha}J_{\alpha'
\alpha}A_{\alpha'}A_\alpha  \label{HC}
\end{equation}
where $J_{\alpha \alpha '}$ takes the value  1  if the $\alpha$ and
$\alpha'$ edges cross each other, and takes the value 0 otherwise. 
Here $\lambda$ and $\gamma$ are non-negative parameters of the model representing the drive for making connections and the crossing penalty strength respectively. In
terms of the Ising spin variables, (\ref{HC}) becomes (apart from an additive constant)
\begin{equation}
{\cal H}=-\sum_{\alpha}\left(\lambda-{{w_\alpha}\over
2}-{{\gamma}\over{N_b}} \sum_{\alpha'} J_{\alpha \alpha
'}\right)S_\alpha+\frac{\gamma}{2 N_b }\sum_{\alpha'\neq \alpha}
J_{\alpha \alpha '}S_{\alpha'} S_\alpha, \label{HC2}
\end{equation}
which is a dilute anti-ferromagnetic Ising system under a local external field $ h_\alpha\equiv  \lambda-{w_\alpha \over 2}-{\gamma\over{N_b}}\sum_{\alpha '}J_{\alpha\alpha'}$.
For further analytical calculations, we take $w_\alpha= d_\alpha +c_o$ and  assume $d_\alpha$'s are drawn from some distance distribution $P_d(d_\alpha)$ (which can be calculated theoretically), and the connection weight distribution is simply given by $P(w)=P_d(w-c_o)$.
%\subsubsection{crude MFT}
%\begin{equation}
%x_o=\tanh\left(\beta[\lambda-\frac{{\bar d}}{2}-(1+\frac{x_o}{2})\gamma {\bar J}]\right)
%\end{equation}
%When $\beta \rightarrow \infty $, we obtain
%\begin{equation} x_o=  \left\{\begin{array}{ll}-1 ,& \mbox{$\lambda <\frac{{\bar d}+\gamma{\bar J}}{2} $}\\
%\frac{2\lambda-{\bar d}}{\gamma{\bar J}} -2     , & \mbox{$
%\frac{{\bar d}+\gamma{\bar J}}{2}\leq \lambda
% \leq \frac{{\bar d}+3\gamma{\bar J}}{2}$}\\1       , & \mbox{$\lambda > \frac{{\bar d}+3\gamma{\bar J}}{2}$}.
%\end{array}\right.
%\end{equation}
%The result of $x_o$ versus $\lambda$ for $\beta\to \infty$ is shown in Figure \ref{crudeMF}.
%\begin{figure}[tbp]
%\subfigure[]{\includegraphics*[width=\figws]{MFcrossxo.eps}}
% \subfigure[]{\includegraphics*[width=\figws]{MFcrossphasediagram.eps}}
%\caption{(a)Mean-field solution of $x_o$ as a function of $\lambda$
%in the $\beta\to \infty$ limit. (b)Mean-field phase diagram of a
%network with edge crossing penalty.} \label{crudeMF}
%\end{figure}

%\subsection{Mean-field Theory (Naive MF)}
Using standard techniques in statistical mechanics, the mean-field equations are obtained as
\begin{equation}
m_\alpha=\tanh\beta \left( h_\alpha-{{\gamma}\over{N_b}} \sum_{\alpha'} J_{\alpha \alpha'}m_{\alpha '}\right)\qquad \alpha=1,2,\cdots, N_b,\label{NMFEq}
\end{equation}
where $\beta$ is the ``inverse temperature'' parameter corresponding to the inverse of intrinsic noise strength for the growth of the optimized network. $m_\alpha=\langle S_\alpha\rangle$ is the average local spin value.  (\ref{NMFEq})  are $N_b$ coupled nonlinear mean-field equations which can be solved by standard numerically routines with a number of initial guess trials to search for possible different solutions, and the one with lowest cost is chosen. 
Here we focus on the zero-temperature limit for the fully optimized network.
In the $\beta\to\infty$ limit, it is more convenient to use the adjacency variable $A_\alpha$ to describe the network connection properties, (\ref{NMFEq}) becomes
\begin{equation}
A_\alpha=\Theta\left( \lambda-{{w_\alpha}\over
2}-{{2\gamma}\over{N_b}} \sum_{\alpha'} J_{\alpha \alpha
'}A_{\alpha '}\right)\label{MFeqT0}
\end{equation}
 where $\Theta$ is the Heaviside step-function. The  solutions of the mean-field equations give the network connection candidates for the optimized network and the numerical results will be presented in Sec.  IV. Before that, we shall impose further approximation to derive some analytic results in next section.
 
 For the fully optimized case of zero-temperature, there is  no edge crossing for sufficiently large  $\gamma$, and the critical $\gamma^*$ can be derived  as follows. For $\gamma>0$, the zero-temperature mean-field equation (\ref{MFeqT0}) can be rewritten as 
\begin{eqnarray}
A_{\alpha}&=&\Theta \left[\frac{N_{b}}{2\gamma}\left(\lambda-\frac{w_{\alpha}}{2}\right)-n_\alpha\right], \quad\hbox{where } 
n_\alpha=\sum_{\alpha^{\prime}}J_{\alpha\alpha^{\prime}}A_{\alpha^{\prime}}  \label{MFAT=0}
\end{eqnarray}
is the number of crossings of edge $\alpha$ whose  possible values  are $0,1,2,\cdots ,N_b$.
Denote $w_{min}$ as the minimal value of the weights for this realization, we have $\frac{N_{b}}{2\gamma}\left(\lambda-\frac{w_{\alpha}}{2}\right)\leqslant \frac{N_{b}}{2\gamma}\left(\lambda-\frac{w_{min}}{2}\right)$.  Hence if 
$\frac{N_{b}}{2\gamma}\left(\lambda-\frac{w_{min}}{2}\right)<  1$, then any edge candidate that has a finite number of crossings ($n_\alpha\geqslant1$) will be impossible to connect (i.e. $A_\alpha=0$). In other words, for
\begin{equation} \gamma > \gamma^*\equiv \frac{N_b}{2}(\lambda-\frac{w_{min}}{2}),
 \label{Nocrossing}   \end{equation} 
 the no edge crossing constraint will be satisfied, and the optimized network will have the same configuration for $\gamma \in (\gamma^*,\infty)$.

\section{Crude Mean-field approximation } %PXing2.f,bisect.f
To proceed  analytically, we made further
approximation by replacing $m_{\alpha '}$ in the last term in
(\ref{NMFEq}) by its mean-field value, namely the order-parameter  $x_o\equiv {1\over N_b}\sum_\alpha m_\alpha$. We call such an approximation the crude mean-field approximation, then $x_o$ can be obtained
from the self-consistent equation
\begin{equation}
x_o={1\over N_b}\sum_\alpha\tanh\left(\beta[\lambda-\frac{{w_\alpha}}{2}-
\gamma(1+x_o)\frac{n_\alpha}{ N_b}]\right),
\end{equation}
where $n_\alpha\equiv\sum_{\alpha'}J_{\alpha\alpha'}$ is the maximal possible crossings of the edge $\alpha$ (which is a non-negative integer). Macroscopic quantities, such as the order parameter, are then obtained by taking the average over the distributions of
$n_\alpha$ and $w_\alpha$. Denote the joint probability of  that
$n_\alpha$ and $w_\alpha$ by $P(n,w)$, one has
\begin{equation}
x_o=\sum_{n=0}^{N_b}\int P(n,w)dw
\tanh[\beta(\lambda-{w\over 2}-\gamma {n\over N_b} (1+x_o))].
  \label{MFbeta}
\end{equation}
 One can also define
${\cal P}_x(n)\equiv \int dw P(n,w)$ and $P_w(w)\equiv \sum_{n=0}^{N_b} P(n,w)$ as the crossing probability and weight distribution respectively.
 For a given
distribution of the nodes on a plane, the distribution $P(n,w)$ can be
obtained by direct sampling. It should be noted that since $w_\alpha=d_\alpha+c_o$ depends on the distance between two nodes on the plane, thus   the number of possible crossings over an edge depends on  its length in general, i.e. $n$ and $d$ (and hence $w$ also) are correlated, namely
$ P(n,w)\neq{\cal P}_x(n) P_w(w)$. This can be directly verified in simulations  by plotting the number of possible crossing for an edges versus its length for nodes randomly placed on a unit square as displayed in Fig. \ref{Pdfig}a in Appendix I, showing that  $n$ and $d$ are in general positively correlated.
${\cal P}_x(n)$ can be obtained from
simulations of $N$ nodes randomly distributed on a square or from $\int dw P(n,w)$ and the results are shown in Fig. \ref{Pdfig}c %and \ref{Pdfig}d 
in  Appendix I.

Fig. \ref{beta}a shows $x_o$ solved from (\ref{MFbeta}) as a function of $\lambda$ together with the Monte Carlo simulation results for $\gamma=1$ and different values of $\beta$. $x_o$ (hence $\langle {\bar A}\rangle$) increases with $\lambda$ and approaches the fully connected limit for sufficiently large $\lambda$.  The crude MF  approximation agrees well with the simulation results at higher temperatures but deviations are  prominent at low temperatures.
\begin{figure}[htbp]
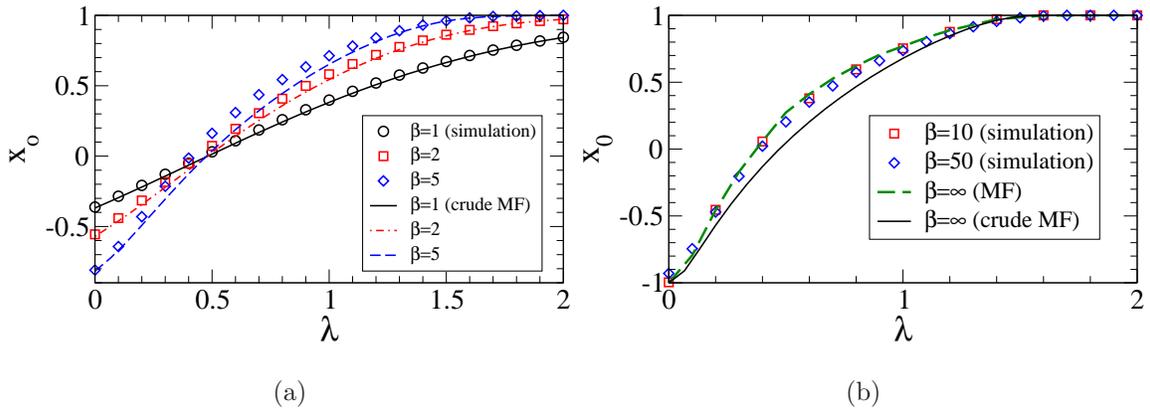

\centering
 \subfigure[]{\includegraphics*[width=\figws]{VNMFxovslambetagam1.eps}}
\subfigure[]{\includegraphics*[width=\figws]{VNMFxovslamT=0gam1.eps}}
\caption{Monte Carlo simulation (symbols) results of the order-parameter $x_o$ vs. $\lambda$ for $N=100$ and $\gamma=1$. (a) For various
values of $\beta$. The crude mean-field approximation  theoretical results (curves)  obtained from (\ref{MFbeta}) are also shown. (b) Monte Carlo simulation results for large 
values of $\beta$. The zero-temperature theoretical results obtained from  crude mean-field approximation (\ref{CMFbarA}) (solid curves) and from the numerical solutions of the mean-field equations (\ref{MFeqT0}) (dashed curve) are also displayed. } \label{beta}
\end{figure}

We then focus on the the fully optimized case of  zero-temperature. In the $\beta\to \infty$ limit, in terms of the average of the mean connection $\langle{\bar A}\rangle$, the crude mean-field equation reads
\begin{eqnarray}
\langle{\bar A}\rangle&=&\sum_{n=0}^{N_b}\int_{-\infty}^{2\lambda-\frac{4\gamma n}{N_b}\langle{\bar A}\rangle} P(n,w)dw\label{CMFbarA0}\\
&=&\sum_{n=0}^{N_b}P^{cum}_w\left(n,2\lambda-\frac{4\gamma n}{N_b}\langle{\bar A}\rangle\right)\label{CMFbarA}
\end{eqnarray}
where $P^{cum}_w(n,u)\equiv \int_{-\infty}^u dw P(n,w)$ is the cumulative probability of $P(n,w)$.
Under the crude mean-field approximation, $\langle \bar{A} \rangle$ can be obtained from the root of Eq. (\ref{CMFbarA}) as a function of $\lambda$ and $\gamma$. 
 Solving  $\langle \bar{A} \rangle$ from Eq. (\ref{CMFbarA}) with $P_w^{cum}(n)$  recorded from direct sampling, the crude mean-field
results for the zero-temperature($\beta \to \infty$) is shown in
Fig. \ref{beta}b for the zero-temperature order parameter $x_o$ (solid curve) as a function of $\lambda$ for $\gamma=1$ together with the Monte Carlo simulation results (symbols) for large values of $\beta$, indicating that the crude mean-field approximation is reasonably accurate, albeit with noticeable deviations from simulation results.  
It is clear that for sufficiently low values of $\gamma$ and large enough $\lambda$, the network is fully connected. 

\begin{figure}[htbp]
\centering
%   \subfigure[]{\includegraphics*[width=\figws]{kvslambdagammaN100.eps}}
%\subfigure[]{\includegraphics*[width=\figws]{kvsgammlambdaN100.eps}}
%{\includegraphics*[width=\figwss,angle=-90] {plot.ps}}
 \subfigure[] {\includegraphics*[width=\figws] {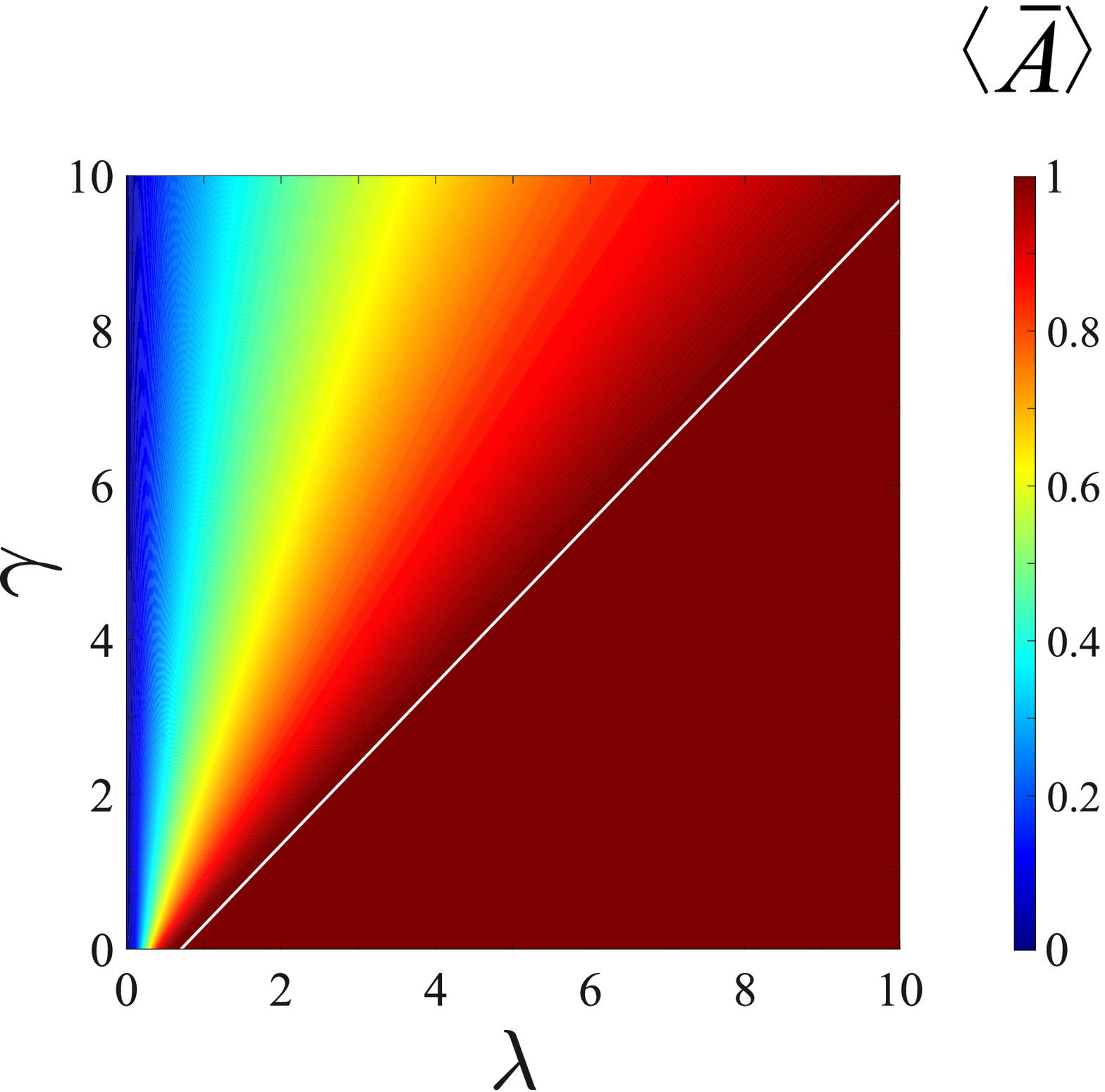}}
  \subfigure[] {\includegraphics*[width=\figws] {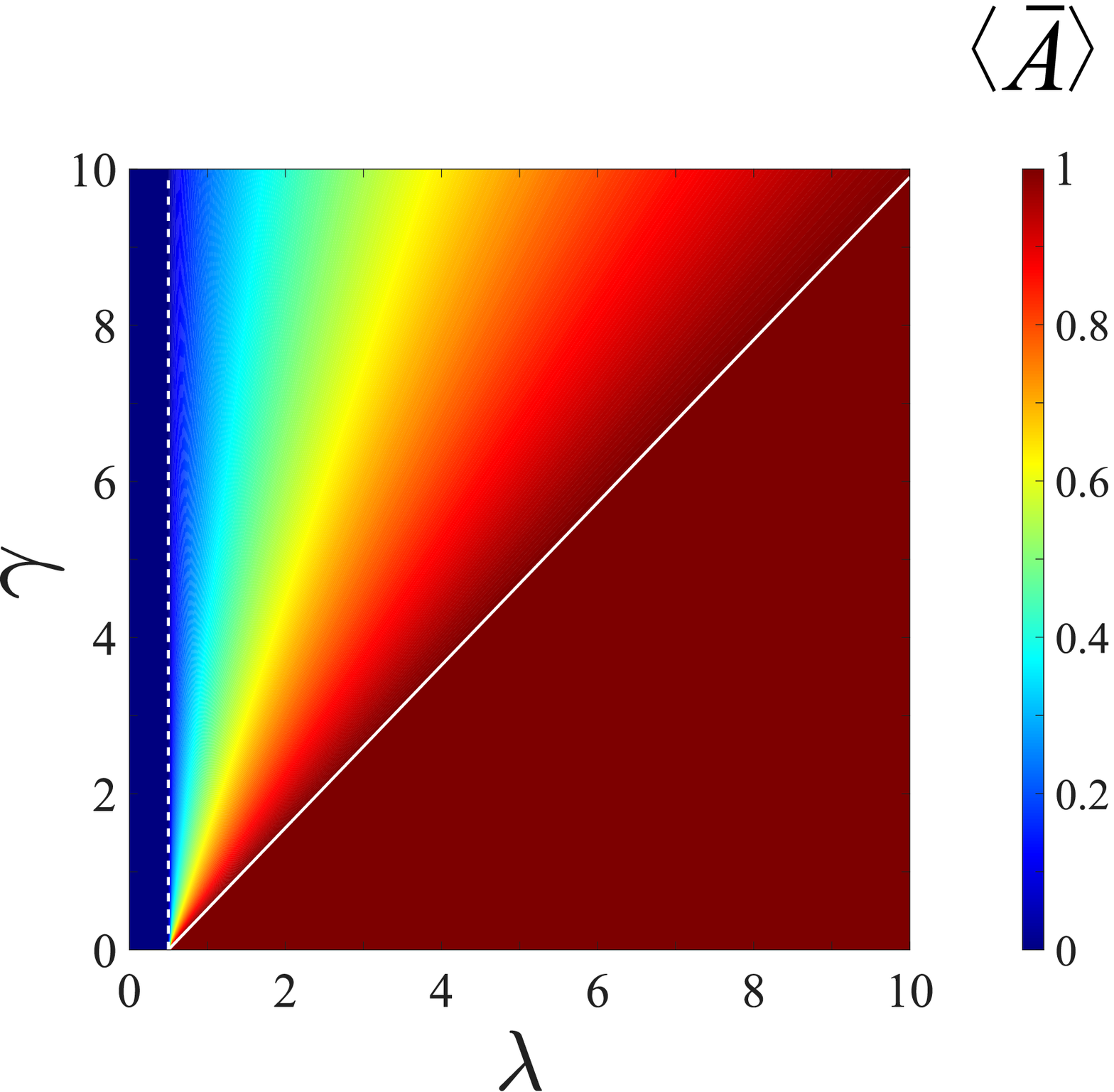}}
  \caption{(a) Phase diagram of the  mean connectivity $\langle \bar{A} \rangle$ of a fully optimized (zero-temperature) two-dimensional network with   bond-crossing cost obtained from the solution of (\ref{CMFbarA}) under the crude Mean-field approximation. $N=100$ and $c_0=0$. 
   The straight line marks the estimated phase  boundary $\lambda \geqslant 0.96\gamma +{1\over \sqrt{2}}$   for the fully connected network with ${\bar A}=1$. (b) Similar to (a) but for the case of $w_\alpha=c_0=1$ which  corresponds to the case in Sec. IIIA. The vertical straight line marks the phase  boundary $\lambda \leqslant  \frac{c_0}{2}$ for the unconnected network with ${\bar A}=0$. The straight line marks the estimated phase  boundary $\lambda \geqslant \frac{c_0}{2}+0.96\gamma$   for the fully connected network with ${\bar A}=1$.
   }
  \label{phasediag}
\end{figure}

Fig. \ref{phasediag} shows the zero-temperature phase
diagram  of $\gamma$ vs. $\lambda$ for the mean connectivity obtained from the root of Eq. (\ref{CMFbarA}).
One can see that for sufficiently large $\lambda$, the optimized network is completely connected and its phase boundary can be deduced as follows.  From (\ref{CMFbarA0}), it is clear that $\langle \bar{A} \rangle=1$ is a root if the upper limit of the integral exceeds the maximal value of the weight, $w_{max}$.
In addition, the maximal number of crossings (denoted by $n_{max}$) can be obtained from sampled cumulative distribution $P_x^{cum}$ under the condition $P_x^{cum}(n_{max})=1$. Hence the condition for $\langle \bar{A} \rangle=1$ is $2\lambda \geqslant 4\gamma n_{max}/N_b+w_{max}$. For instance with $N=100$, from the sampled $P_x^{cum}(n)$ (see Fig. \ref{Pdfig}d), one gets $n_{max}/N\simeq 0.48$ and for nodes randomly distributed on a unit square $w_{max}=d_{max}=\sqrt{2}$ (for $c_0=0$), one obtains the phase boundary line $\lambda = 0.96\gamma +\frac{1}{\sqrt{2}}$. This theoretical phase boundary line is also plotted in Fig. \ref{phasediag} showing very good agreement.

\begin{figure}[htbp]
\centering
 \subfigure[]{\includegraphics*[width=\figwss]{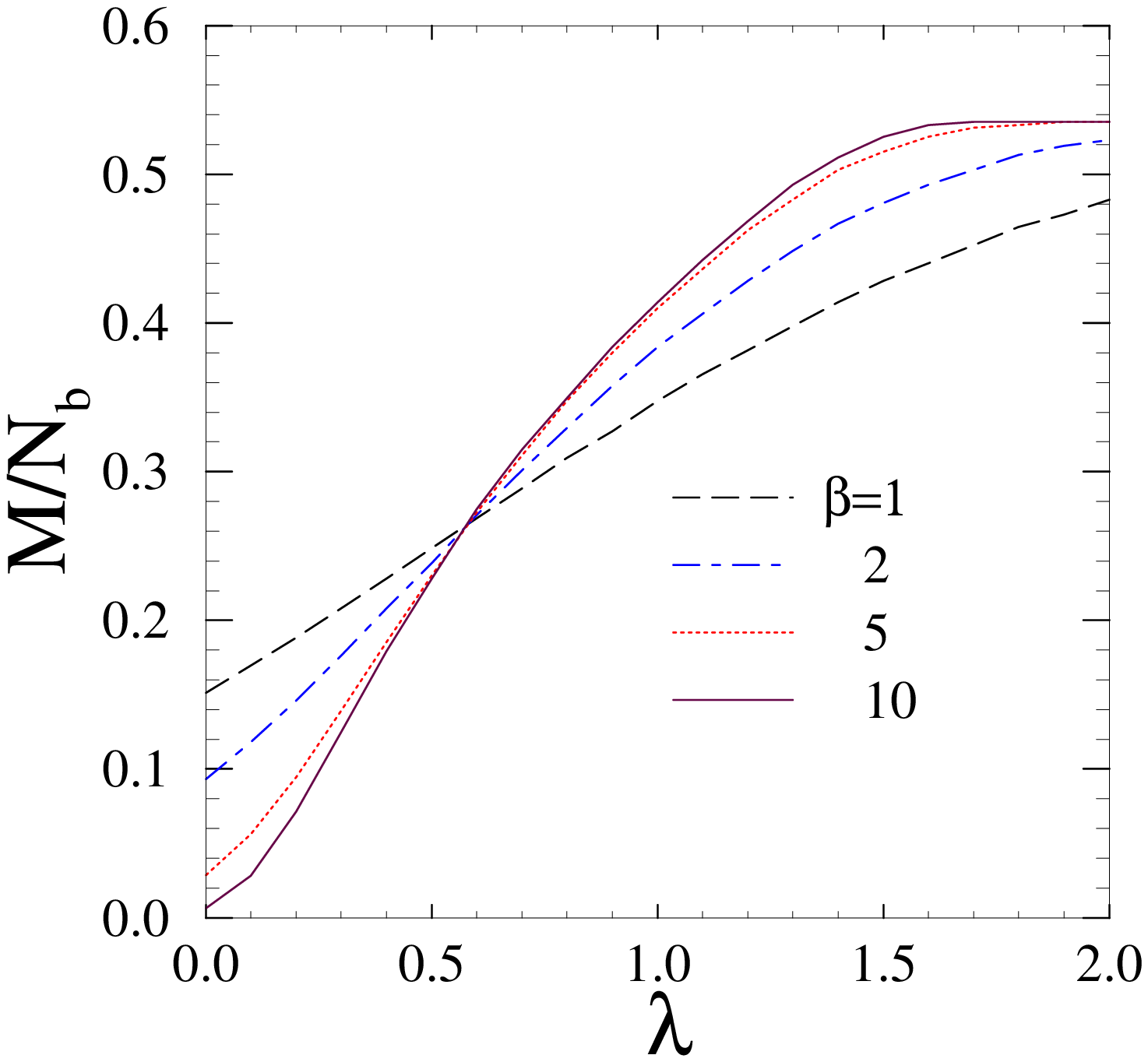}}
 \subfigure[]{\includegraphics*[width=\figwss]{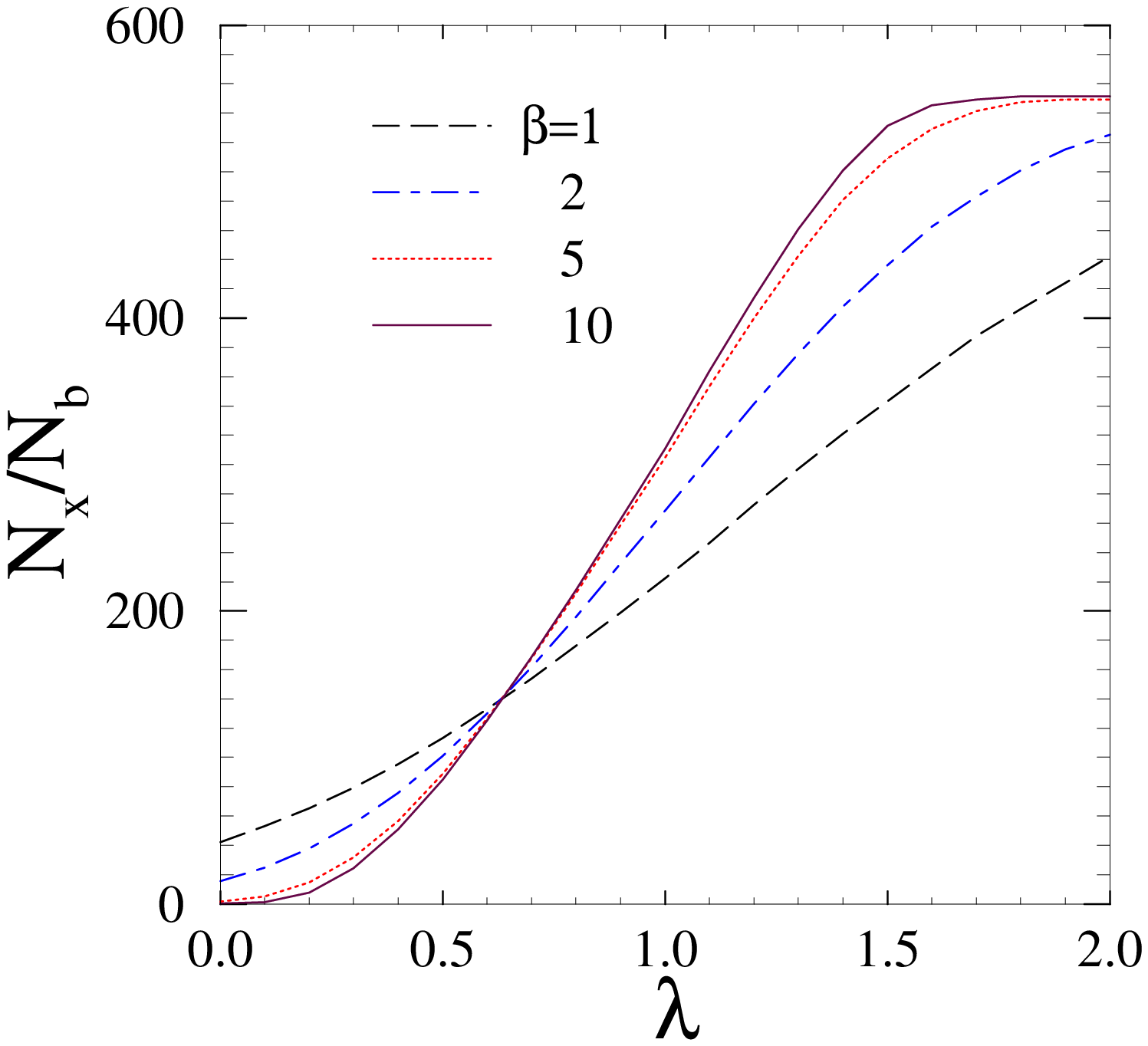}}
 \subfigure[]{\includegraphics*[width=\figwss]{ModCMNxvskgam1_.eps}}
\caption{Monte Carlo simulation results  of the (a) material cost $M$
and (b) mean crossing number as a function of $\lambda$ for
different values of $\beta$ with $\gamma=1$.  (c)  Log-log plot of
the data in (a) and (b) versus $\langle {\bar A} \rangle$ to verify the scaling prediction from the crude mean-field approximation. The
dotted and  solid lines are of slopes 1 and 2 respectively.}
\label{ModCMNx}
\end{figure}

%If one assumes that $w_\alpha$ and $n_\alpha$ are independent, then $ P(n,w)={\cal P}_x(n) P_w(w)$. For illustration purposes,  if $P_w(w)$ is taken to be uniformly distributed in [0,1],  integrating (\ref{MFbeta}) gives \begin{equation}x_o={2\over\beta}\sum_{n=0}^{N_b} {\cal P}_x(n) \ln \left(\frac{\cosh[\beta(\lambda-\gamma {n\over N_b} (1+x_o))]}{\cosh[\beta(\lambda-{1\over 2}-\gamma {n\over N_b} (1+x_o))]}\right).\label{finitebeta}\end{equation}
The average material cost $M\equiv\langle \sum_\alpha d_\alpha
A_\alpha\rangle$ and number of crossing $N_x\equiv
\langle\sum_{\alpha'<\alpha}J_{\alpha'
\alpha}A_{\alpha'}A_\alpha\rangle$ are also measured from simulations.
Fig. \ref{ModCMNx}a and  \ref{ModCMNx}b show the Monte Carlo results of the average material cost and number of crossings as a function of $\lambda$ with $\gamma=1$ for several values of $\beta$. Both $M$ and $N_x$ increases with $\lambda$, and  more sharply at low temperature. 
Under the  crude mean-field approximation, it is easy to see that $M\simeq
\left(\sum_\alpha
d_\alpha\right) {\langle {\bar A}\rangle }$ and $N_x \simeq
\left(\sum_{\alpha'<\alpha}J_{\alpha'
\alpha}\right)\langle {\bar A}\rangle ^2$, for
different values of $\beta$,  $\lambda$ and $\gamma$. Fig. \ref{ModCMNx}c plots the Monte Carlo simulation results of $M$ and $N_x$ vs. $\langle {\bar A}\rangle$ verifying  the scaling  of $M$ and $N_x$ with $\langle {\bar A}\rangle$ of the crude mean-field predictions.

\subsection{case of $c_0\gg d_\alpha$}
\begin{figure}[htbp]
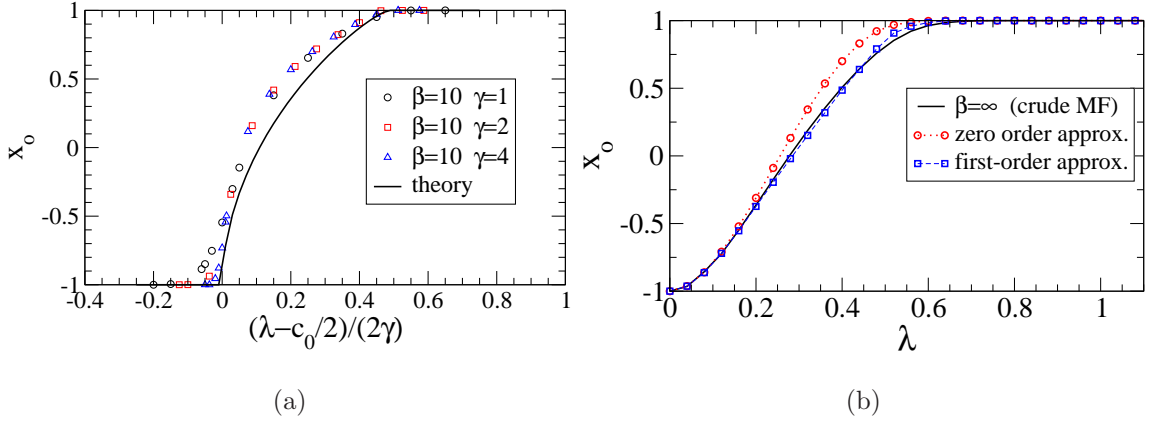

 \centering
\subfigure[]{\includegraphics*[width=\figws]{xoc0gamlam.eps}}
 \subfigure[]{\includegraphics*[width=\figws]{CMFxovslamsmallgam_1.eps}}
\caption{(a) The case of $c_0 \gg d_\alpha$:  Monte Carlo simulation results for the case of $w_\alpha=c_0$  showing $x_o$ vs. the scaled parameter $\frac{\lambda-c_0/2}{2\gamma}$ for various values of $\gamma$ at low temperature $\beta=10$. The crude mean-field  theoretical approximation of (\ref{CMFbarAc0}) at zero temperature is also shown (curve).  (b) Crude mean-field  approximation result of $x_o$ vs. $\lambda$ for $\gamma=0.1$ (curve). Explicit approximations of zero and first order calculated from (\ref{0-order}) and (\ref{1st-order}) (symbols) respectively are also  shown to verify the analytic formulas for small values of $\gamma$. }\label{CMFfig}
\end{figure}

On the other hand, the correlation between $n_\alpha$ and $w_\alpha$ can be negligible under some special condition. For instance if the node connection cost outweighs the wiring cost, i.e. $c_0\gg d_\alpha$, then $w_\alpha\simeq c_0$ leading to the independence between $n_\alpha$ and $w_\alpha$: $ P(n,w)={\cal P}_x(n) P_{c_0}(w)$, where $ P_{c_0}$ is the distribution function of $c_0$. In this case the zero-temperature crude mean-field equation (\ref{CMFbarA}) becomes
 \begin{eqnarray}
\langle{\bar A}\rangle&=&\sum_{n=0}^{N_b}{\cal P}_x(n)  \int_{0}^{2\lambda-\frac{4\gamma n}{N_b}\langle{\bar A}\rangle}  P_{c_0}(u)du\\
&=&\sum_{n=0}^{N_b} {\cal P}_x(n) \Theta\left(\lambda-\frac{c_0}{2}-\frac{2\gamma n}{N_b}\langle{\bar A}\rangle\right)\quad \hbox{for }  P_{c_0}(u)=\delta(u-c_0)\label{CMFbarAc0theta}\\
&=& {\cal P}_x^{cum}(n^*)\label{CMFbarAc0}
\end{eqnarray}
 where the case of a fixed $c_0$ is considered in above, and ${\cal P}_x^{cum}(m)\equiv \sum_{n=0}^{m}{\cal P}_x(n) $ is the cumulative probability of the crossing of the possible edges.  $n^*$ is given by the largest integer  that $n^*{\cal P}_x^{cum}(n^*)$ is smaller than $\frac{N_b}{2\gamma }(\lambda-\frac{c_0}{2})$. It is clear for this case that $\langle{\bar A}\rangle$ is a function of a single scaled parameter $\frac{\lambda-\frac{c_0}{2}}{2\gamma }$. This can be verified from the Monte Carlo simulation results of $x_o$ at low temperatures for various values of $\gamma$ and $\lambda$ which all collapse onto a master curve close to the theoretical curve from the zero-temperature crude mean-field approximation as shown in Fig. \ref{CMFfig}a.
 The phase diagram  for the mean connectivity in  this case obtained from the root of Eq. (\ref{CMFbarA}) is shown in Fig. \ref{phasediag}b.
 It is easy to see from (\ref{CMFbarAc0theta}) that $\langle{\bar A}\rangle=0$ (unconnected network) is the solution if $\lambda \leqslant  \frac{c_0}{2}$. Also   from (\ref{CMFbarAc0}) that $\langle{\bar A}\rangle=1$ (fully connected network) is the solution for $\lambda \geqslant \frac{c_0}{2}+2\gamma \frac{n_{max}}{N_b}$. These two phase boundaries are also plotted in  Fig. \ref{phasediag}b showing nice agreement.
 
 \subsection{case of small $\gamma$}
Obviously, there will be no edge crossing penalty if $\gamma=0$.
% Below we will show that the edge penalty cost is negligible already for sufficiently small (but still finite) $\gamma$. 
Consider the zero-temperature crude mean-field equation (\ref{CMFbarA0}) under the condition of $\gamma \ll \lambda/2$, since $\frac{n}{N_b}<1$ and $\langle{\bar A}\rangle \leq 1$, one can expand the rhs of  (\ref{CMFbarA0}) for small $\gamma$ to give
\begin{eqnarray}
\langle{\bar A}\rangle&=& { P}^{cum}_d(2\lambda-c_0)-\frac{2\gamma \langle{\bar A}\rangle}{\lambda} g(2\lambda) +{\cal O}((\dfrac{\gamma}{\lambda})^2)\\
&=&\left(1-  \frac{2\gamma }{\lambda} g(2\lambda)\right){P}^{cum}_d(2\lambda-c_0) + {\cal O}((\dfrac{\gamma}{\lambda})^2)\\
&  &\hbox{where } g(x)\equiv \dfrac{x}{N_b} \sum_{n=0}^{N_b} n P(n,x)\label{1st-order}
\end{eqnarray}
 which can be computed from the sampled $P(n,w)$, and is shown in Fig. \ref{Pdfig}e in Appendix I.
 Thus to lowest order, one has
 \begin{equation}
\langle{\bar A}\rangle\simeq { P}^{cum}_d(2\lambda-c_0) + {\cal O}(\dfrac{\gamma}{\lambda}),\label{0-order}
 \end{equation}
 which essential says that the crossing penalty is negligible and the cost reduces to Model A in \cite{cheng}.

  Furthermore, $P_d(d)$ can be calculated analytically in some case, otherwise it can be measured
once the position of the nodes are given. 
For example, if the $N$ nodes are uniform and randomly fall on a unit square, $P_d(d)$ can be calculated (see Appendix I for derivation) to be
\begin{equation}
P_d(X) = \left\{
\begin{array}{ll}
2X(\pi-4X+X^2)     , & \mbox{$0<X\leqslant 1$}\\
2X\left\lbrace 2[\csc^{-1}X-\tan^{-1}\sqrt{X^2-1}-1] +4\sqrt{X^2-1}-X^2\right\rbrace ,& \mbox{$1<X<\sqrt{2} $}.
\end{array}\right.\label{Pd}
\end{equation}
  The analytic results of (\ref{Pd}) together with the measured $P_d(d)$ from simulations are shown in  Fig. \ref{Pdfig}f  in Appendix I .  The cumulative distance distribution, $P_d^{cum}(z)\equiv \int_0^zP_d(u) du$, can be evaluated to give
  \begin{equation}
P_d^{cum}(z)= \left\{
\begin{array}{ll}
z^2(\pi-{8\over 3}z+{z^2 \over 2})    , & \mbox{$0<z\leqslant 1$}\\
 {1\over 3}+ {4\over 3}\sqrt{z^2-1} (2z^2+1)+ 2z^2[\csc^{-1}z-\tan^{-1}\sqrt{z^2-1}-{z^2\over 4}-1],& \mbox{$1<z\leqslant \sqrt{2} $}.\end{array}\right.\label{Iz}
 \end{equation}
 Using the explicit result for $P_d$ in (\ref{Pd}) for nodes that are uniformly distributed randomly in a square domain, Fig. \ref{CMFfig}b shows the order-parameter as a function of $\lambda$ obtained from the root of (\ref{CMFbarA}) together with  the zero and first-order results calculated using (\ref{0-order}) and (\ref{1st-order}), indicating that the 1st-order expansion gives very accurate agreement.

%\subsubsection{network properties}

%\subsubsection{zero-temperature limit}

\subsection{No bond crossing constraint} 
\begin{figure}[htbp]
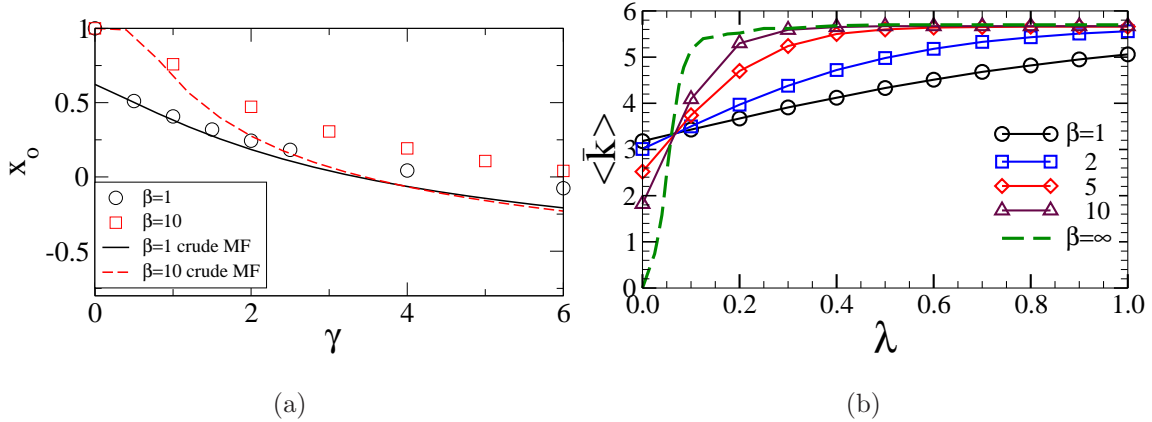

\centering
  \subfigure[]{\includegraphics*[width=\figws]{xovsgamlam1beta.eps}}
     \subfigure[]{\includegraphics*[width=\figws]{ModCkvslamNBXnew.eps}}
%  \subfigure[]{\includegraphics*[width=\figws]{ModCNoXingkvslam.eps}}
  \caption{Monte Carlo simulation results of $x_o$ (symbols) as a function of $\gamma$ for $\lambda=1$ and $\beta=1$ and 10. $N=100$. The corresponding theoretical results of crude mean-field approximation are also shown (curves). (b) Monte Carlo simulation results of the mean degree  $\langle{\bar k}\rangle$ (symbols) as a function of $\lambda$ for various values of  $\beta$, for optimized network under no bond-crossing constraint. The result for the fully optimized network (zero-temperature) using our algorithm in Sec. IV is also shown (dashed curve).  }\label{xovsgam}
  \end{figure}

%NN2DC.f, netsub.f; NN2DB2.f (MC simulations); NaiveMFT=0.f; sorting.f
For fixed values of $\lambda$ and $\beta$, the connectivity of the optimized network decreases as the crossing penalty $\gamma$ increases, as shown by the Monte Carlo simulation and crude mean-field results in Fig. \ref{xovsgam}a. For small values of $\gamma$, the crude mean-field results agrees well with the simulations but deviation becomes more significant as $\gamma$ increases. The total number of crossing of the resultant network decreases with $\gamma$ .

For the case of strictly no bond-crossing is allowed, it is anticipated that the number of links will be suppressed, and the optimization would be strongly frustrated due to the strong interaction for the crossings between the edge candidates. The strictly no bond crossing constraint corresponds to the strongly frustrated anti-ferromagnetic spin system. One may expect there are many solutions with near costs in the low-temperature regime.
To gain some insights, the network is first studied by Monte Carlo simulations under the constraint that there
is strictly no bond crossing. Starting with $\beta=1$, the network
is annealed down to $\beta=10$ slowly. Since under the no bond-crossing constraint, the connectivity of the optimized network is low and hence it would be better to examine the mean degree ${\bar k}$ directly. Fig. \ref{xovsgam}b shows the variation of average mean degree  as a
function of $\lambda$  as the temperature is lowered from $\beta=1$ to $\beta=10$. ${\bar k}$ increases with $\lambda$ and saturates for large values of $\lambda$. The saturation to the maximal value of $\langle {\bar k}\rangle$ occurs for smaller $\lambda$ for lower temperatures and at zero-temperature $\langle{\bar k}\rangle\simeq 5.7$ for $\lambda \gtrsim 0.2$.

To proceed analytically, for the fully optimized network in the zero-temperature limit, the no bond crossing constraint can be considered by taking the $\gamma\to\infty$ limit. Using the crude mean-field approximation, the saddle-point equation (\ref{MFbeta}) gives
\begin{equation}
x_o+1={\cal P}_x(0)+\int P(0,u)\tanh\beta(\lambda-{u\over 2}) du.\label{xoNBX}
\end{equation}
Further take the $\beta\to\infty$ limit in (\ref{xoNBX}) for fully optimized network, one gets
\begin{equation}
\frac{x_o+1}{2}=\langle {\bar A}\rangle =P_w^{cum}(0,2\lambda).
\end{equation}
For nodes that are uniformly random distributed in a unit square, $ P_w^{cum}(0,2\lambda)$ is recorded by direct sampling. Thus the average mean degree is given by $\langle {\bar k}\rangle=(N-1)P_w^{cum}(0,2\lambda)$. This implies that for sufficiently large $\lambda$,  $\langle {\bar k}\rangle=(N-1){\cal P}_x(0)$. Using the measured value of ${\cal P}_x(0)$ for $N=100$, the above result gives $\langle k \rangle \simeq 0.44$ which is much less than that  obtained from Monte Carlo simulations of $\langle {\bar k}\rangle\simeq 5.7$ (see Fig. \ref{xovsgam}b). Such a discrepancy can be understood as follows: 
 the crude mean-field approximation by taking $A_\alpha'$ to be a constant in (\ref{MFeqT0}) in the large $\gamma$ limit will suppress the $A_\alpha$ links for $J_{\alpha\alpha '}=1$ even if $A_{\alpha'}$ vanishes and hence will underestimate $k$. Therefore, for networks obeying the strict no edge crossing constraint, i.e. genuine planar networks, one needs to abandon the crude mean-field approximation and look for the solutions of the system of  mean-field equations which is considered in next section.

%\section{Mean-field Equations for Optimized 2D Network with crossing penalty} % NaiveMFT=0.f; NaiveMFTemp.f, zscnt.f 

\section{Solution of the Mean-field equations, Ground-state Cost Landscape and  Optimal  Algorithm for no-crossing solution  }
 In this section, we shall solve the zero-temperature mean-field equations (\ref{MFeqT0}) numerically and will show that the minimal cost solution indeed gives very accurate results for the order-parameter measured from Monte Carlo simulations at low temperatures.
 In addition, the solution of the mean-field equation provides the precise network connections for the fully optimized network, which is essential for  practical applications. 
The mean-field equations  (\ref{MFeqT0}) are $N_b$ coupled nonlinear equations with discrete binary variables $A_\alpha=0$ or 1.  One can attempt to solve Eqs. (\ref{MFeqT0})  iteratively, i.e. starting with some initial ${\vec A}$ and evaluate an improved ${\vec A}'$ using the rhs of (\ref{MFeqT0}) and repeat the process until it converges to the solution. To gain more insight, consider the component $A_\alpha$ is changed to $A_\alpha '$ upon one iteration.
From (\ref{HC}), one gets the corresponding change in the cost, $\Delta {\cal C}_\alpha/\Delta A_\alpha=-2\lambda+d_\alpha+\frac{4\gamma}{N_b}\sum_{\alpha'}J_{\alpha\alpha '}A_{\alpha '}$, and  $\Delta {\cal C}_\alpha=0$ if $\Delta A_\alpha=0$. Thus $\Delta {\cal C}_\alpha<0$ for  $\Delta A_\alpha\neq 0$ upon iterating (\ref{MFeqT0}), which suggests that the cost will be decreased upon the iteration dynamics of connecting or disconnecting a single link. In practice, the mean-field equations are solved iteratively with a large number of initially trials and attempt to exhaust all the solutions, then the solution with the lowest cost is chosen. Fig. \ref{beta}b displays the results  of the order-parameter $x_o$ (related to the mean connectivity via (\ref{meank})), obtained from solution of the mean-field equations (dashed curve), as a function of $\lambda$ for $\gamma=1$, showing significant improvement over the crude mean-field approximation and excellent agreement with the low temperature Monte Carlo simulation results. However, in some situations, especially for larger $\gamma$, attempt to find the solution iteratively or by other standard system of nonlinear equations solver may not converge in affordable computational times.

As shown Fig. \ref{beta}b, the crude mean-field approximation at $\beta\to \infty$ can predict the gross variation of the order-parameter with $\lambda$. However, for large values of $\gamma$,  deviations  become rather significant. Fig. \ref{xovsgam}a plots the Monte Carlo simulation results together with the crude mean-field results as a function of $\gamma$ for $\lambda=1$, showing  increasing deviations as $\gamma$ becomes larger even for $\beta=1$. In addition, numerical solutions  of the full mean-field equations for the lowest cost solution become inefficient as $\gamma$ becomes large and in practice very difficult to find the mean-field solution(s) unless $N$ is small ($\leq 10)$.
%The solutions are also shown in Fig. \ref{beta}a for comparison. For low values  of $\beta$, the crude mean-field results agree well with the full mean-field solution, but the full mean-field results becomes more accurate for higher values of $\beta$. ??

\begin{figure}[htbp]
\centering
 \subfigure[]{\includegraphics*[width=\figws]{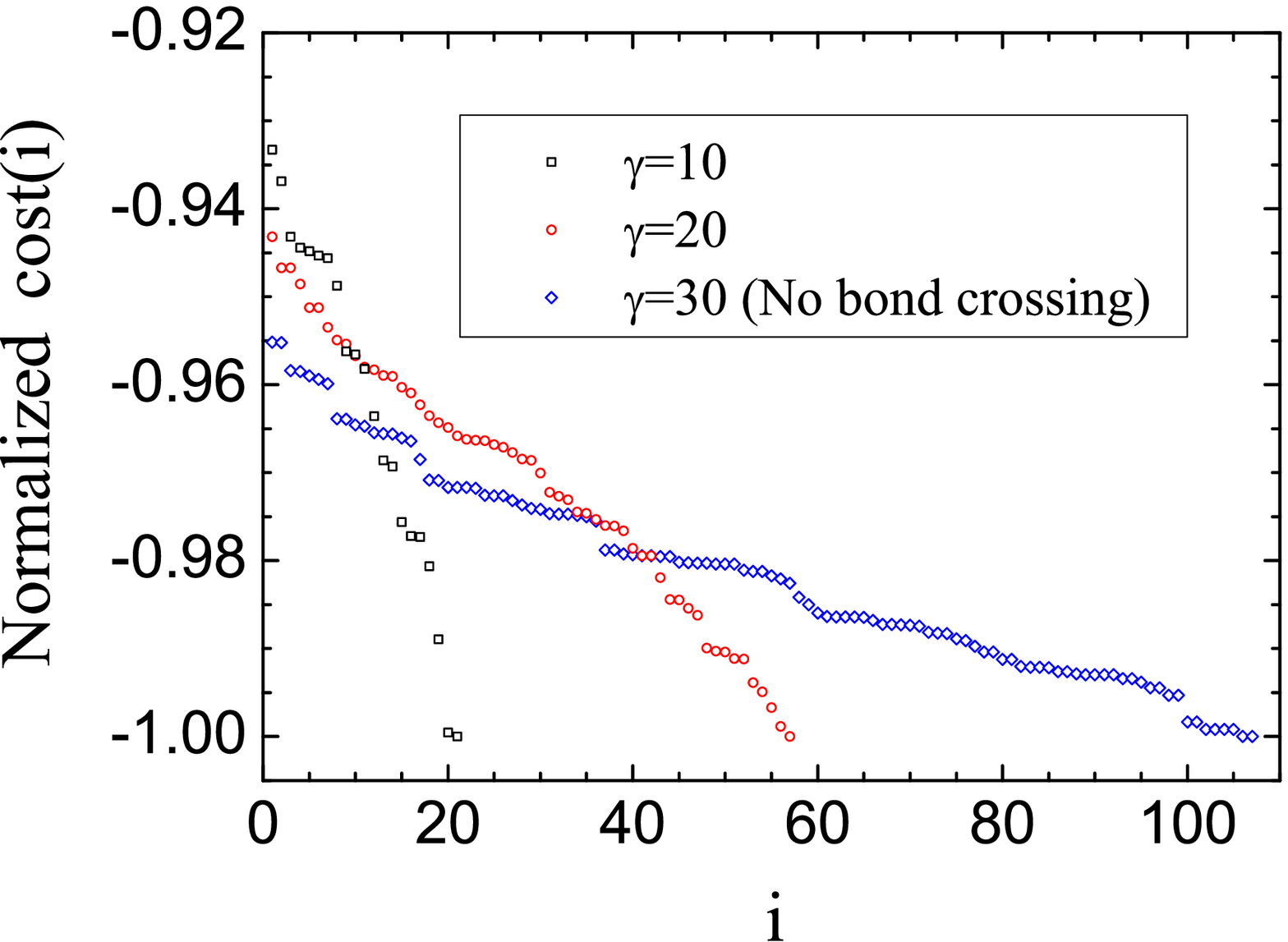}}
\subfigure[]{\includegraphics*[width=\figws]{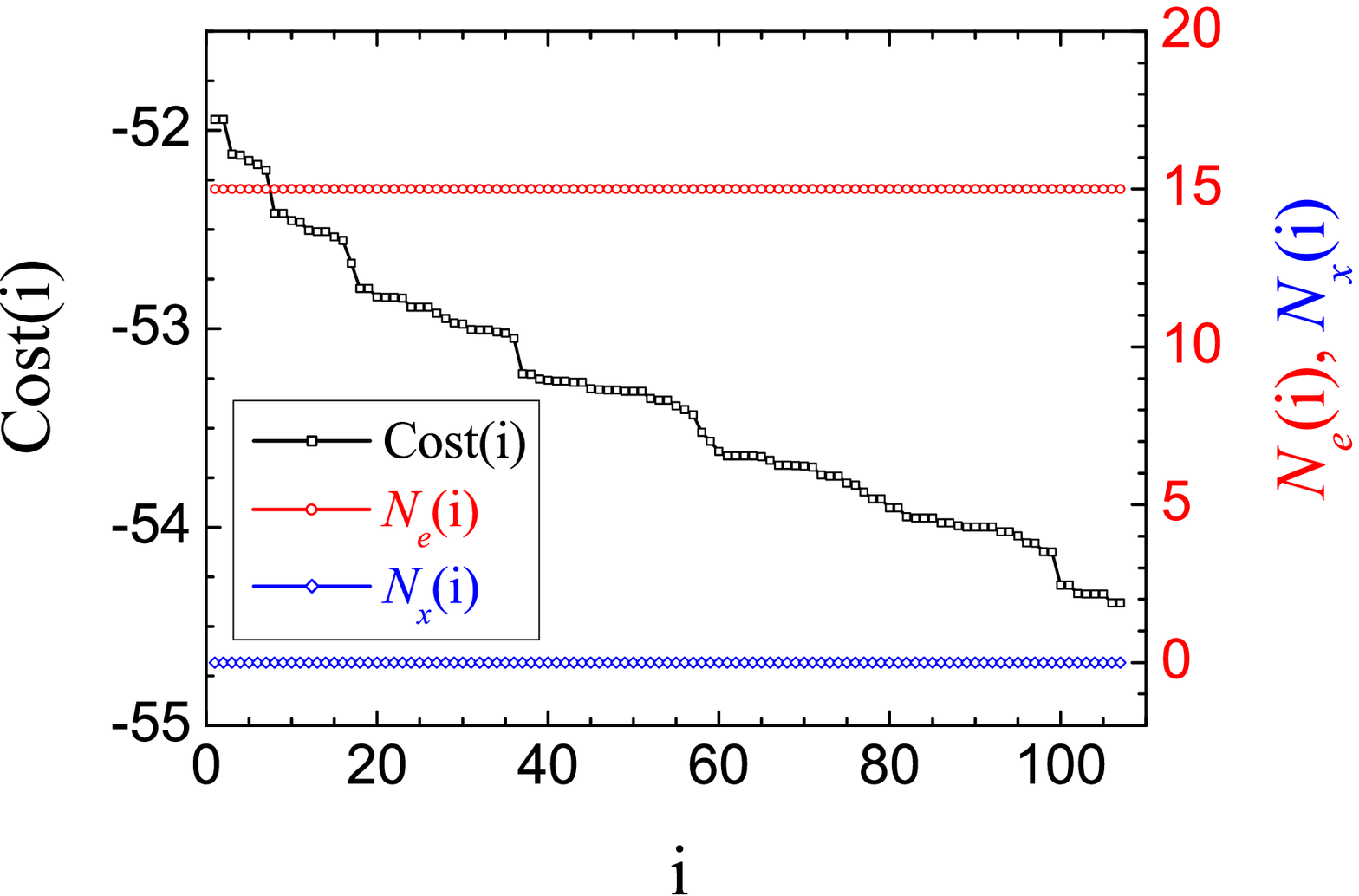}}
% \subfigure[]{\includegraphics*[width=\figwss]{Figure1costiN10.eps}} 
\subfigure[]{\includegraphics*[width=\figws]{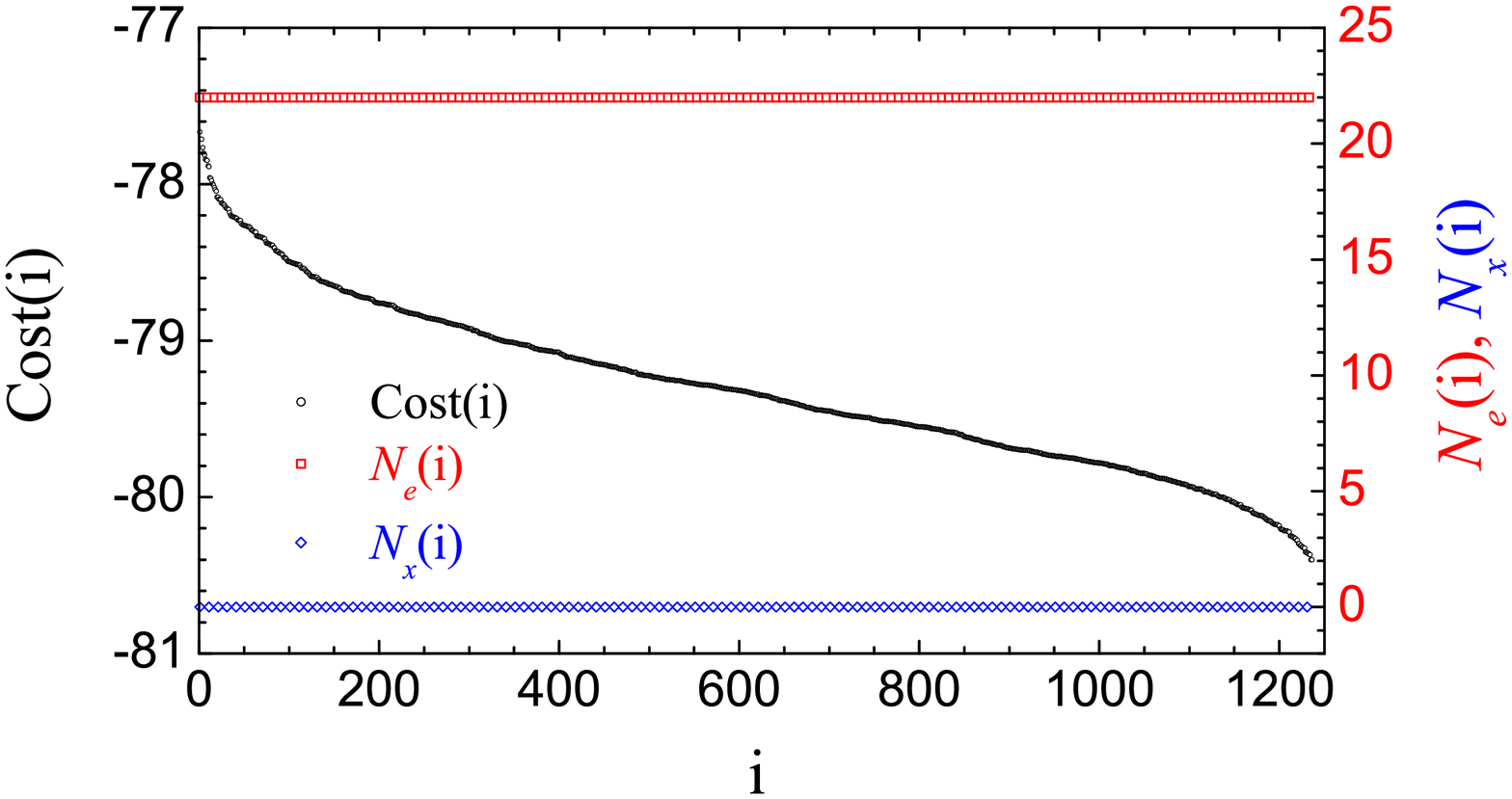}} 
\subfigure[]{\includegraphics*[width=\figws]{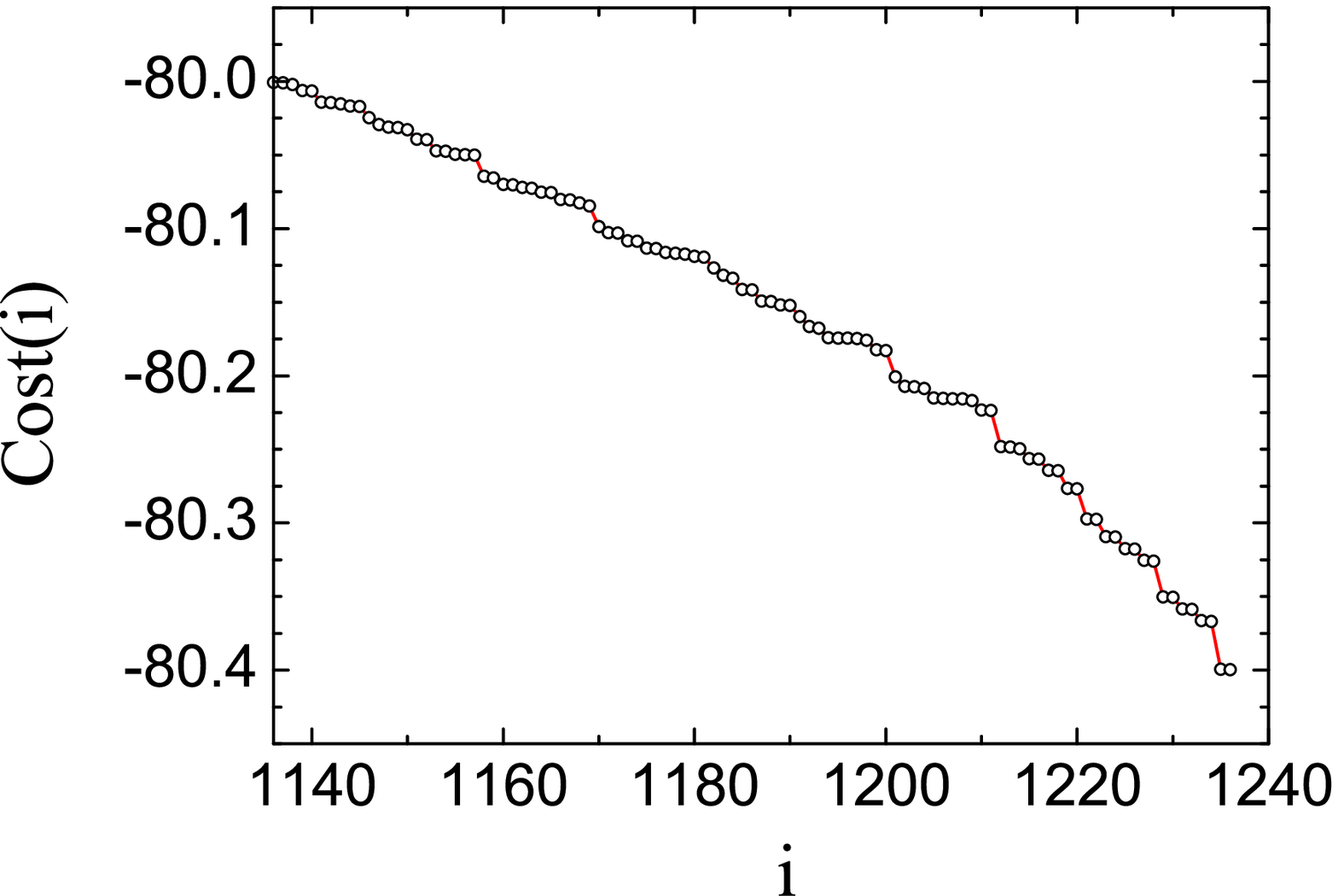}}
  \caption{Cost landscape of all mean field solutions  obtained by enumeration for various bond-crossing penalty   and  $\lambda=2$.  (a)   The normalized cost landscape for $N=8$ with  $\gamma=10,$ 20, 30 for the same realization of the node positions. In this case, $\gamma=30$ exceeds the critical $\gamma^*$ and the resultant network satisfy the no bond crossing condition.  (b) $N=8$ ($N_b=28$) with $\gamma=30$(large enough so that there is no bond-crossing). The number of edges ($N_e$) and crossings ($N_x$) are also shown on the right vertical axis. The number of crossings is zero. Notice that the number of edges for all distinct solutions are identical. (c)   Similar to (a) but for $N=10$ ($N_b=45$) and a sufficiently large  $\gamma=100>\gamma^*$ that no bond-crossing is allowed. In this realization,  there are altogether 1236 mean-field solutions with indices arranged in decreasing order of cost. (d) A zoom-in near the regime of lowest cost solutions to show more clearly the complex rugged landscape.}  \label{landscape}
\end{figure}

\begin{figure}[htbp]
\centering
%%{NMFSvslambda.eps}}
%%{NMFSvsgamma.eps}}
%%{DeltaEFvsgamma.eps}}
%%{DeltaEgvsgamma.eps}}
\subfigure[]{\includegraphics*[width=\figws]{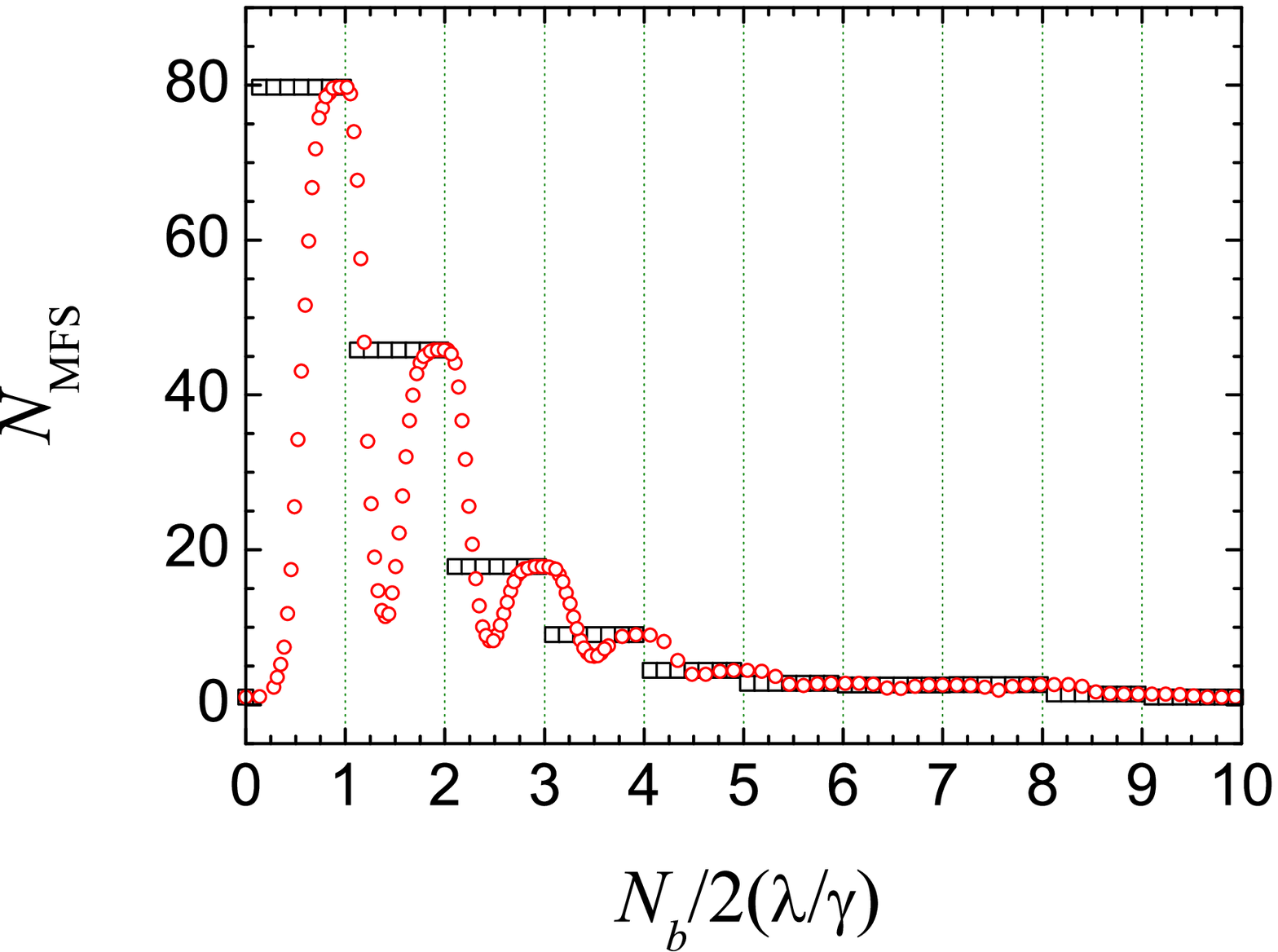}} \subfigure[]{\includegraphics*[width=\figws]{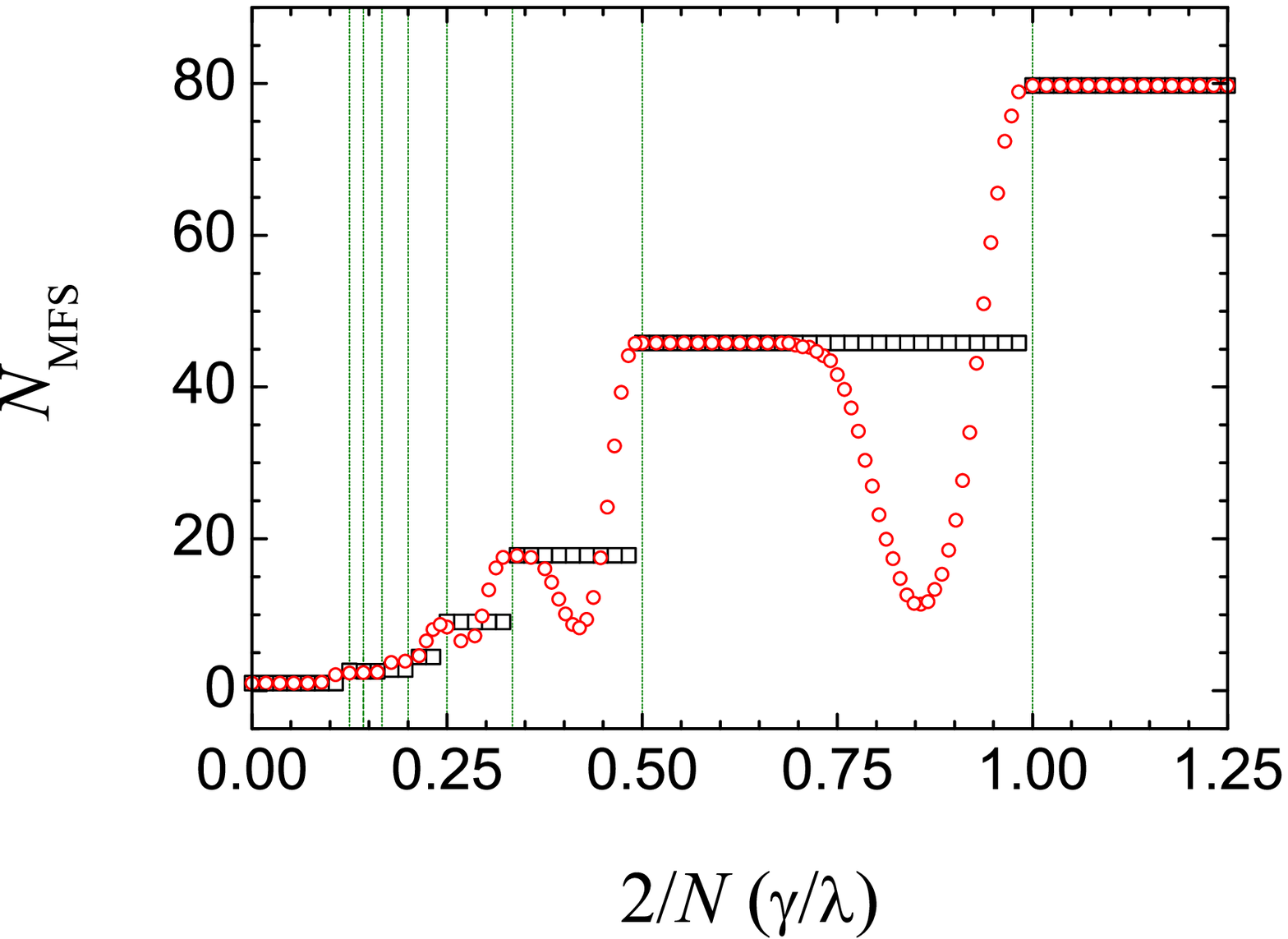}}
\subfigure[]{\includegraphics*[width=\figws]{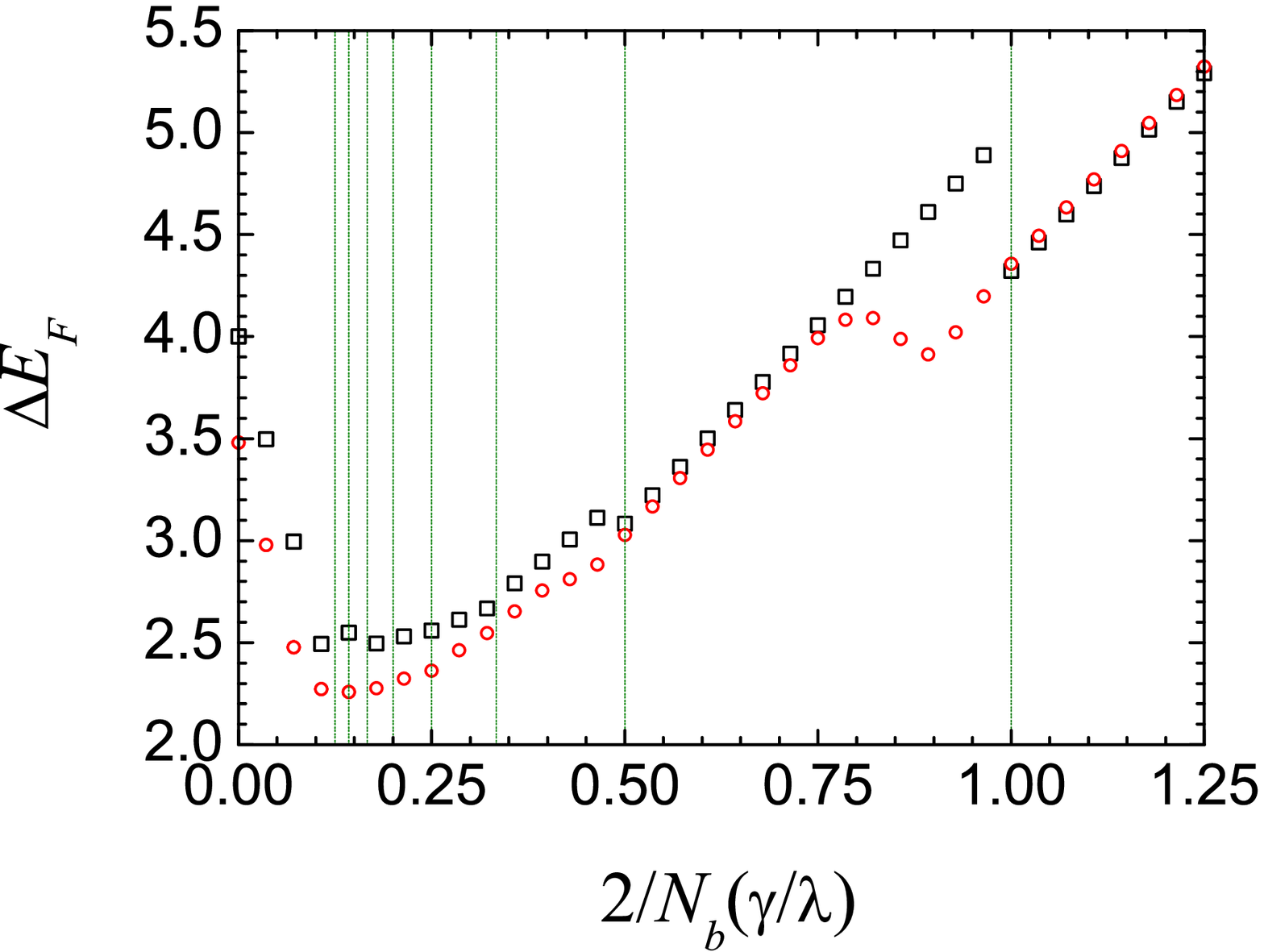}}
\subfigure[]{\includegraphics*[width=\figws]{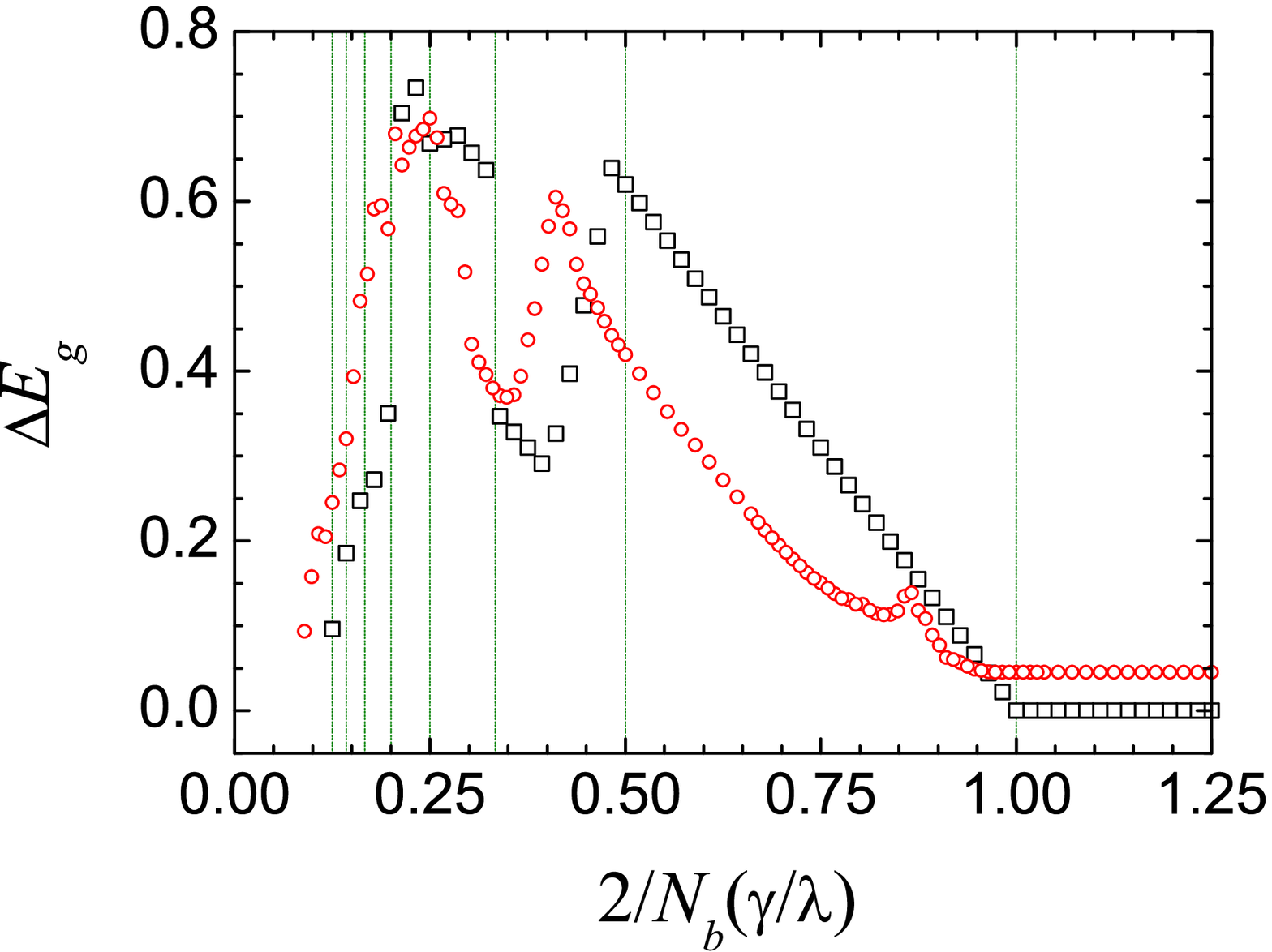}}
  \caption{Analysis of the solutions of the mean-field equations  by enumeration for $N=8$, as a function of the scaled parameter $\frac{N_b}{2\gamma}(\lambda-\frac{c_0}{2})$ or its inverse. $c_0=0$ is taken without loss of generality.   $\square$ refers to the simplified case of zero edge weights,i.e. $w_\alpha\equiv0$, and $\circ$ corresponds to the usual case of the edge cost is the distance between the two nodes, $w_\alpha=d_\alpha$. An average of 200 realizations of randomly placed node location is performed. (a) The number of mean-field solution, $N_{\rm MFS}$ vs. $\frac{N_b}{2\gamma}(\lambda-\frac{c_0}{2})$. The vertical lines mark the integer values of 1,2,3 $\cdots$. (b)  The results in (a) with $N_{\rm MFS}$ plotted against  $\frac{2\gamma}{N_b(\lambda-\frac{c_0}{2})}$ showing the increasing trend of $N_{MFS}$ with $\gamma$. The vertical lines mark the fractional values of 1,$\frac{1}{2}, \frac{1}{3}, \cdots$.  (c) The average cost or energy change due to the flip of a bit in the mean-field solution, $\Delta E_F$ vs.  $\frac{2\gamma}{N_b(\lambda-\frac{c_0}{2})}$.  (d)   The cost difference or energy gap between the ground-state(lowest cost) first excited state (second lowest cost) mean-field solutions, $\Delta E_g$ vs.  $\frac{2\gamma}{N_b(\lambda-\frac{c_0}{2})}$. }
  \label{DeltaEFEg}
\end{figure}

As $\gamma$ increases, the anti-ferromagnetic interaction leads to strong frustrations, resulting in a highly complex ground state energy landscape. This is echoed by the  decrease in the number of convergent solutions obtained by iteration as the landscape  gets more rugged.
For the case of strictly no bond crossing, 
 attempt to find the solutions of the mean-field equations (\ref{MFeqT0}) for $\gamma>\gamma^*$ by iterations  can lead to period-2 orbits and will not converge unless the initial trial hits the solution by chance (but with a low probability that decays exponentially with $N_b$).
Nevertheless, one can still gain further insights by examining the cost landscape in detail for small $N$.
 For small values of $N$, one can enumerate all possible zero-temperature mean-field solutions and examine their costs to gain some insight of the cost landscape of the system.  Fig. \ref{landscape}a plots the cost of all the mean-field solutions for $N=8$  in descending order of the cost , for a certain realization of nodes, and for several large values of $\gamma$.  The number of mean-field solutions increase with $\gamma$ and reaches a maximal number of 107 when $\gamma>\gamma^*$. The cost landscape becomes more rugged and the cost spacings are more closely packed as $\gamma$ increases and the system becomes more frustrated. As shown in Fig. \ref{landscape}b, the number of crossings vanishes when $\gamma>\gamma^*$. In addition, the total number of edges $N_e$ (and hence the mean degree) are identical for all these mean-field solutions  under the no bond-crossing constraint, which can be proved theoretically later on.
 Similar behavior is observed for $N=10$ (also by enumeration) under the no bond-crossing constraint, as shown in Fig. \ref{landscape}c.
The cost landscape indicates that the solutions can have  very close energies, as shown in \ref{landscape}d more clearly under magnification.
To gain further insights on the nature of the complex cost landscape due to  frustration  as the crossing penalty increases, we examine how the number of mean-field solutions ($N_{\rm MFS}$) changes with $\gamma$, computed by enumeration for $N=8$. It would be  more intuitive to first consider the simplified case of a constant edge weight, i.e. $w_\alpha =c_0$, ($c_0$  can be taken to be 0 without loss of generality) as in Sec. IIIA. As shown in Fig. \ref{DeltaEFEg}a, $N_{\rm MFS}$ shows a stepwise decrease whenever $\frac{N_b}{2\gamma}(\lambda-\frac{c_0}{2})$ passes an integer value. This can be understood from (\ref{MFAT=0}) with $w_\alpha=c_0$, since $n_\alpha$ are non-negative integers, rhs of (\ref{MFAT=0}) remains the same  when $\frac{N_b}{2\gamma}(\lambda-\frac{c_0}{2})$ lies between two consecutive integers. More precisely, for $m\leqslant \frac{N_b}{2\gamma}(\lambda-\frac{c_0}{2}) <  m+1$ ($m$ is a non-negative integer) the number of crossings in the  network is at most $m$.  For the general case of $w_\alpha=d_\alpha+c_0$,  $N_{\rm MFS}$ shows modulations due to the variations in the distances between two nodes. For $m\leqslant \frac{N_b}{2\gamma}(\lambda-\frac{c_0}{2}) <  m+1$, a fraction of the edges has at most $m$ crossings while the remaining portion of the edges have at most $m+1$ crossings. Fig. \ref{DeltaEFEg}b plots $N_{\rm MFS}$ vs.  $\frac{2\gamma}{N_b(\lambda-\frac{c_0}{2})}$ showing the stepwise increase in $N_{\rm MFS}$ as frustration becomes stronger, agreeing with the idea that the cost landscape is filled with more local minima leading to the difficulty in searching for the global minimal cost in such a complex rugged landscape.

We then further look into the landscape structure by perturbing each mean-field solution by flipping one bit in ${\vec A}$ (i.e by deleting or adding one edge) a time and probe the cost/energy changes. As shown in the beginning of this section, the energy always increases when any one bit of any mean-field solution is flipped indicating that each mean-field solution possesses the characteristics of a local energy minimum in this high-dimensional cost landscape. Denote the energy change of flipping one bit of the mean-field solution averaged over $N_b$ bits and also averaged over all the mean-field solutions by $\Delta E_F$, Fig. \ref{DeltaEFEg}c plots $\Delta E_F$ as a function of $\frac{2\gamma}{N_b(\lambda-\frac{c_0}{2})}$ for the cases of $w_\alpha=c_0$ and $w_\alpha=d_\alpha+c_0$. For larger values of $\gamma$,  $\Delta E_F$ shows a general increasing trend with  $\gamma$, agreeing with the anticipation that the average local minimum well depth increases with frustration.
We also examine the first energy gap,  $\Delta E_g$ defined as the cost gap between the global minimal cost and the second lowest cost of the mean-field solutions. Fig. \ref{DeltaEFEg}d plots $\Delta E_g$ as  function of $\frac{2\gamma}{N_b(\lambda-\frac{c_0}{2})}$ for the cases of $w_\alpha=c_0$ and $w_\alpha=d_\alpha+c_0$. For the case of $w_\alpha=c_0$, , $\Delta E_g=0$  for $\gamma>\gamma^*$ (no bond crossing), indicating that all mean-field solutions are degenerate with the cost $C=(c_0-2\lambda)N_e$. For edge dependent weights, such as  $w_\alpha=d_\alpha+c_0$, $\Delta E_g$ shows a general decreasing trend with $\gamma$ for $\frac{2\gamma}{N_b(\lambda-\frac{c_0}{2})}\gtrsim 0.5$, indicating the low energy states have energy level more closely spaced as frustration grows, resulting in more complex landscape.

Instead of trying to find the solution of (\ref{MFeqT0}) by brute-force method such as enumeration or random initial guesses(which can only work for small $N$), we proposed a new way to find the (near) optimal solution of (\ref{MFeqT0}), outlined as follows. We first resort back to the original mean-field  equations and look for other more practical algorithm for the strictly no bond crossing  case. Taking the $\gamma\to \infty$ limit, Eq. (\ref{MFeqT0}) reduces to
\begin{equation}
A_\alpha=\Theta\left( \lambda-\dfrac{w_\alpha}{2}
\right) \left\lbrace 1-\Theta(n_\alpha)\right\rbrace.\label{MFeqT0NBX}
\end{equation}
We first show that the total number of edges ($N_e$) of different networks satisfying (\ref{MFeqT0NBX}) have the same value, which was verified in Fig. \ref{landscape}b and  \ref{landscape}c. Using (\ref{MFeqT0NBX}), one gets $N_e=\sum_\alpha A_\alpha=\sum_{n_\alpha=0 \And w_\alpha<2\lambda}$, i.e. $N_e$ is simply obtained by counting the number of edge candidates that satisfy $n_\alpha=0$ and $w_\alpha<2\lambda$, which is a fixed integer for give nodes and weights realization.  The optimized cost is simply $C_{opt}=\sum_\alpha (w_\alpha-2\lambda)A_\alpha=\sum_\alpha w_\alpha A_\alpha-2\lambda N_e$.

\begin{figure}[htbp]\centering
 % \subfigure[]{\includegraphics*[width=\figws]{CostLandscapeN=10gam1lam_5.eps}} \subfigure[]{\includegraphics*[width=\figws]{CostLandscapeN=10gam1lam_5.eps}}
 %  \subfigure[]{\includegraphics*[width=\figwss]{CostLandscapeN=10gam4lam_5.eps}}   \subfigure[]{\includegraphics*[width=\figwss]{CostLandscapeN=10gam8lam_5.eps}}\subfigure[]{\includegraphics*[width=\figwss]{CostLanscapeN=10NBX.eps}}\subfigure[]{\includegraphics*[width=\figwss]{CostLanscapeN=10NBXzoom.eps}}\subfigure[]{\includegraphics*[width=\figwss]{hitrateN10lam_5.eps}}\subfigure[]
  {\includegraphics*[width=\figws]{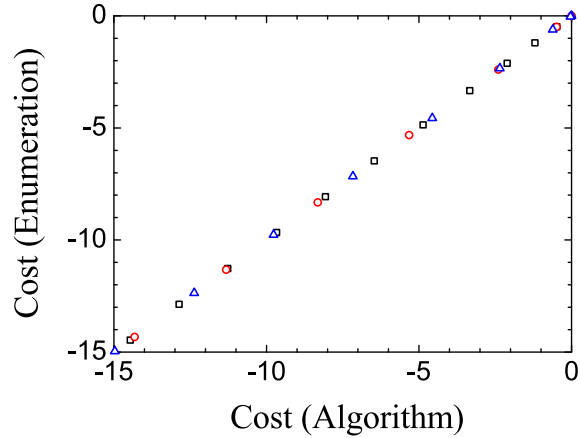}}
  \caption{%(a) Cost Landscape  with strictly no bond-crossing constraints with $N=10$ and large values of $\lambda$. The width of the platform reflects the relative size of the attractive basic of the solution with the corresponding cost. 
  Minimal cost of the optimized network with no bond-crossing found by enumeration  for $N=8$  vs.  the minimal cost obtained by our algorithm, for 3 different realizations (denoted by different symbols) of the nodes on the unit square,  with various values of $\lambda$.  }
  \label{costcost}
\end{figure}

Motivated by the no bond crossing mean-field equation  (\ref{MFeqT0NBX}), here we propose an efficient algorithm of finding the (near) lowest cost solution  as follows:
With a fully connected network (all $A_\alpha$'s=1),  starting with the edge of minimal $w$ (say $w_\sigma$), remove this connection if 
 $w_\sigma > 2\lambda$, otherwise for those $J_{\sigma\alpha'}=1$ put $A_\alpha'=0$. Repeat the above process for the next minimal weight edge till all edges are gone through once.
Although there is no guarantee that the solution obtained by the above algorithm is the ground state solution, but it will always give a solution much much faster than solving the coupled nonlinear equations and has no problem of convergence.
Fig.\ref{costcost}  shows the minimal cost found by enumeration  for $N=8$ ($N_b=28$ and $2^{N_b}= 268 435 456$) compared with the cost obtained by our method, for 3 different realizations of the nodes on the unit square,  with various values of $\lambda$. The results clearly verify  that our method finds the minimal cost in all cases, but with a computational time several orders of magnitude faster.
In addition, in all the cases we studied, we found that the solution obtained by our efficient algorithm always has a lower cost as compared to that found by solving the coupled nonlinear equations or by Monte Carlo simulations annealed to low temperatures.

\section{Optimized Network properties}
\begin{figure}[htbp]
\centering
  \subfigure[]{\includegraphics*[width=\figwss]{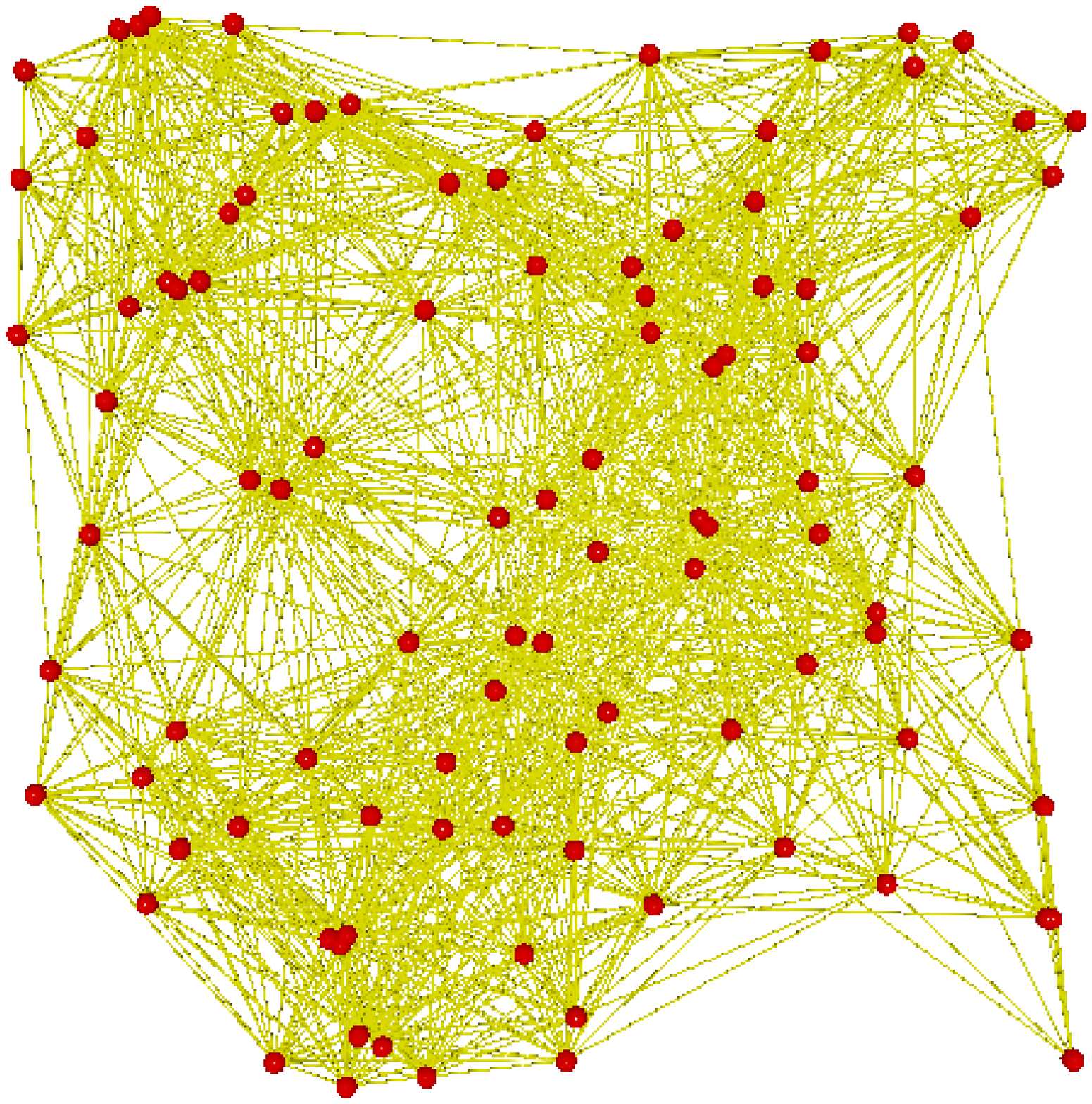}}
   \subfigure[]{\includegraphics*[width=\figwss]{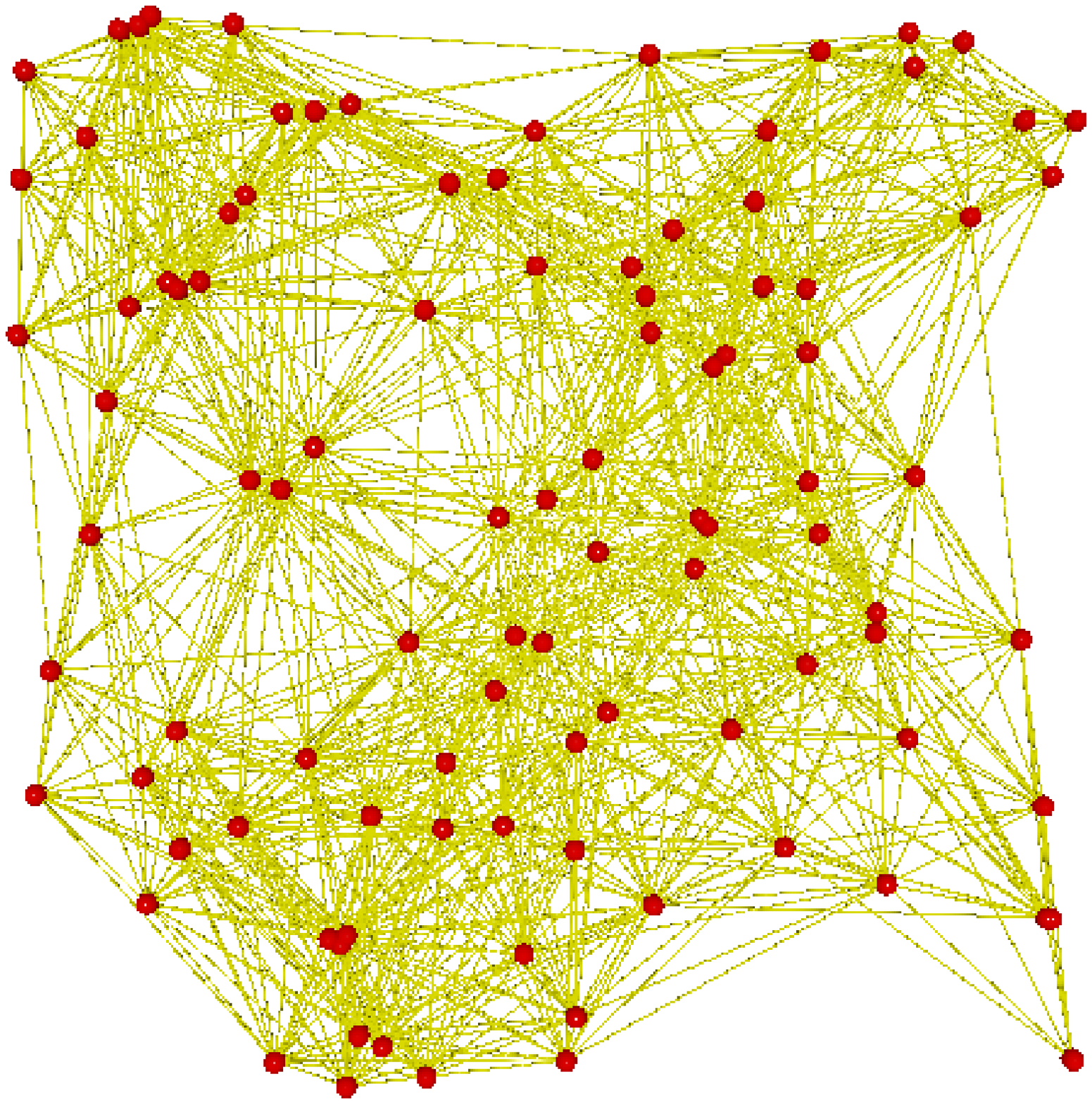}}
   \subfigure[]{\includegraphics*[width=\figwss]{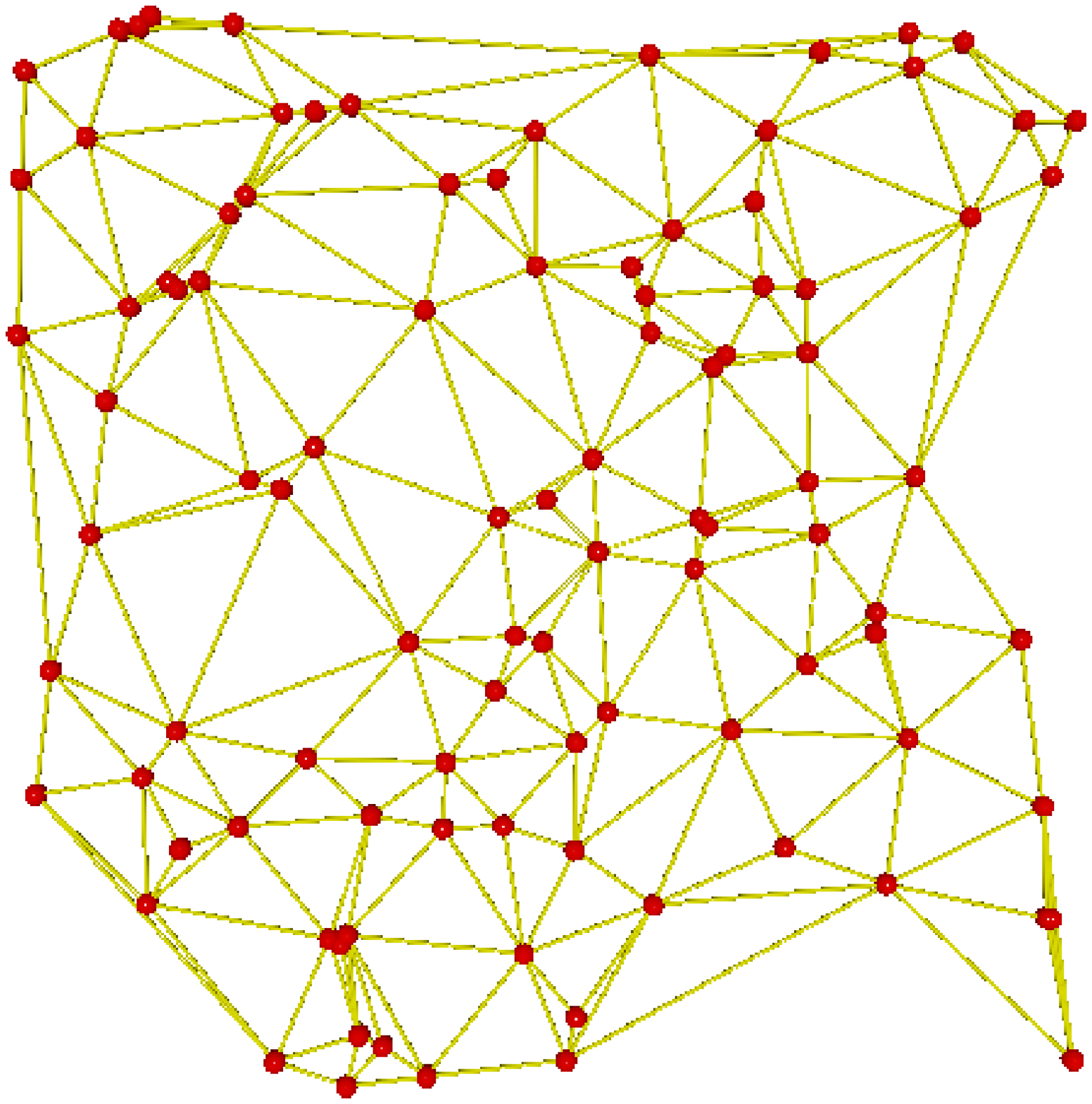}}
  \caption{Results of the fully optimized (zero-temperature)  network configurations  with $N=100$ and $\lambda=0.2$ obtained from the solution of the mean-field equations for (a) $\gamma=0.2$, (b) $\gamma=1$, and (c) strictly no bond-crossing ($\gamma=\infty$), using the optimal algorithm  in Sec. IV. }
  \label{NMFT=0lam_2pov}
\end{figure}

\begin{figure}[htbp]
\centering
   \subfigure[]{\includegraphics*[width=\figwss]{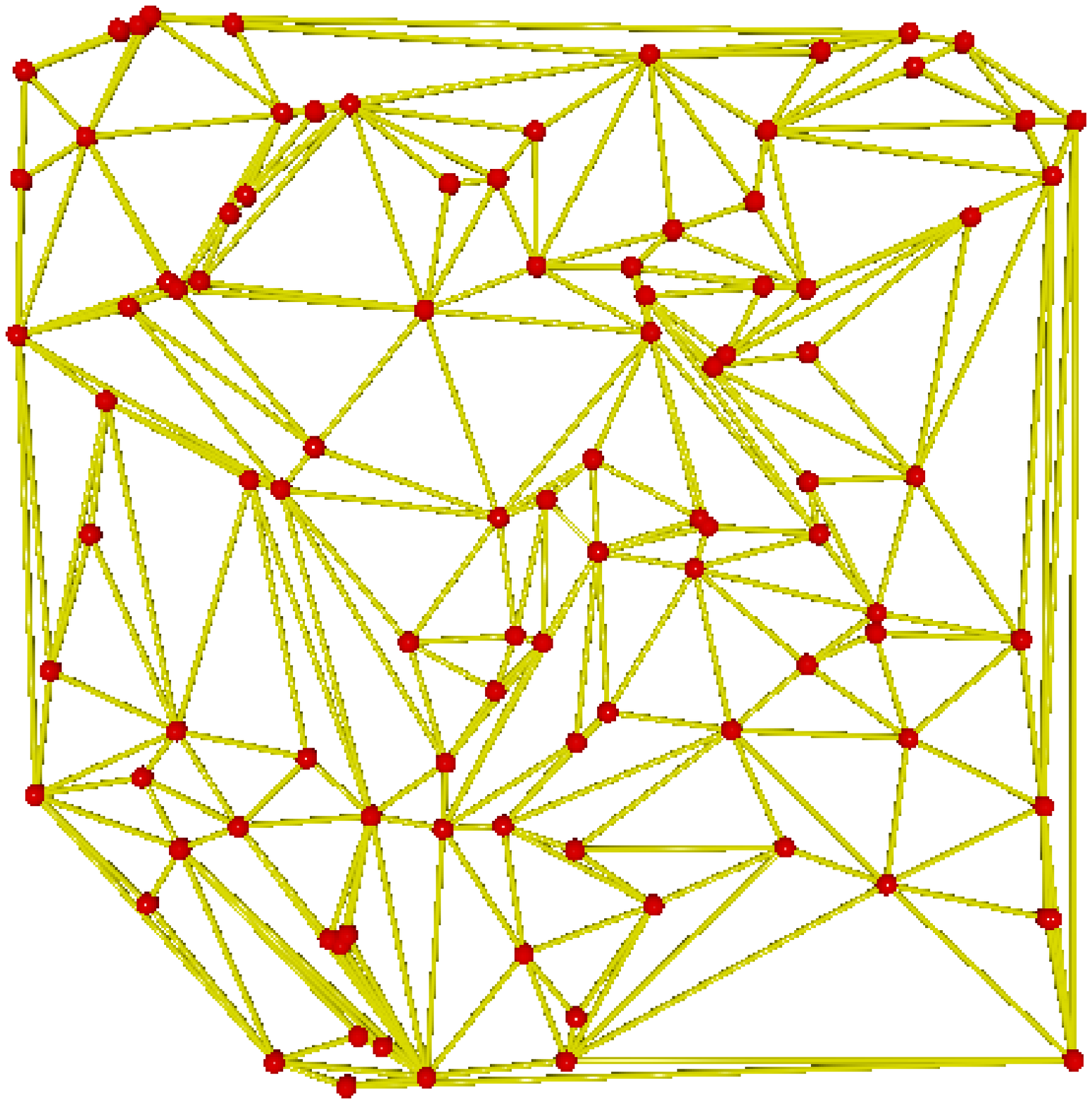}}  \subfigure[]{\includegraphics*[width=\figwss]{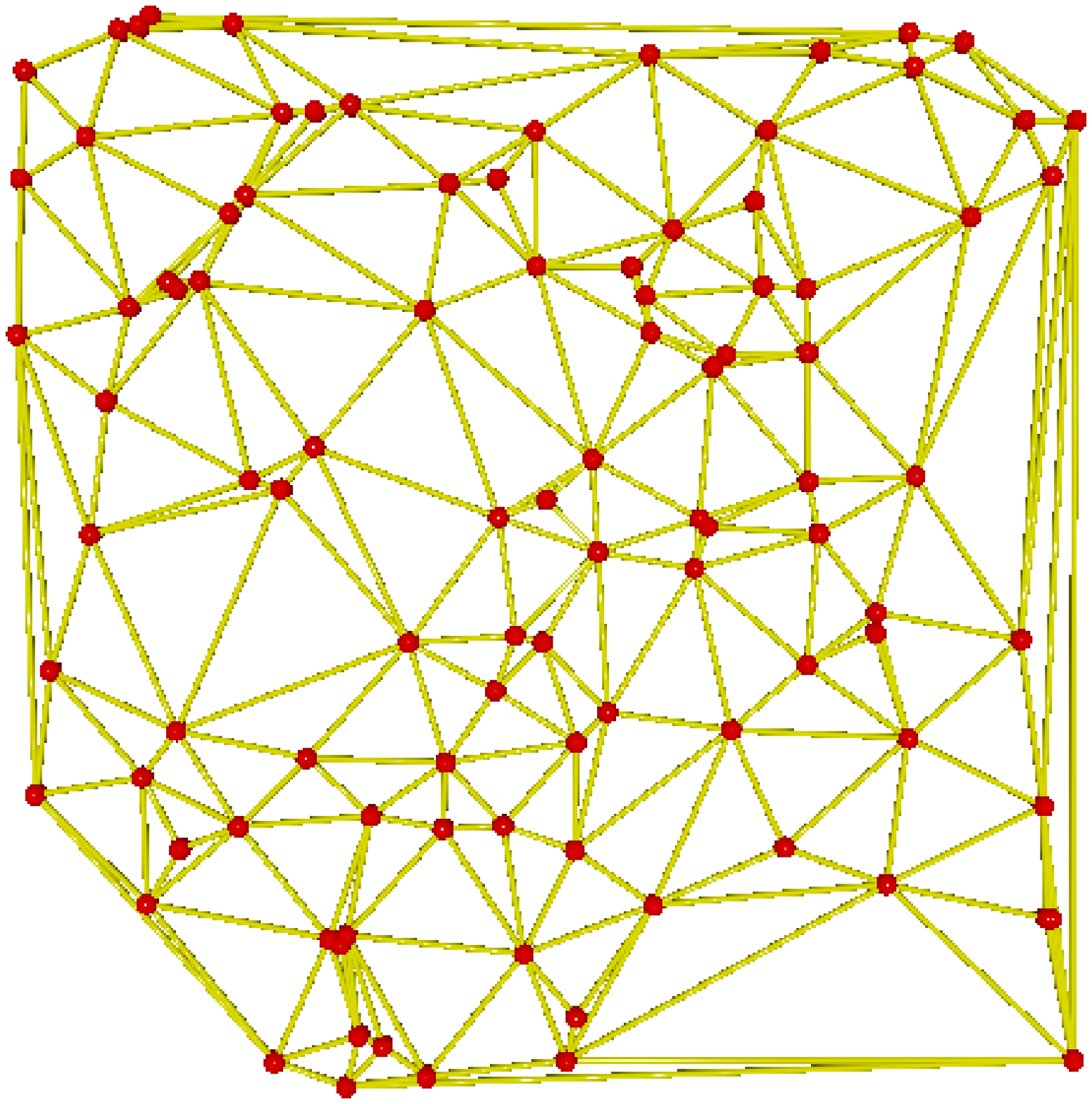}}
      \subfigure[]{\includegraphics*[width=\figwss]{Pstarlam1NBX.eps}}
%        \subfigure[]{\includegraphics*[width=\figws]{BX2DPkscale.eps}}
  \caption{ Results of the  optimized  planar 2D  network configurations with no bond-crossing constraint. $N=100$ and $\lambda=1$ (a) Monte Carlo simulation result annealed down to $\beta=50$. Cost $=-527.735$.  %=42.265 - 570.
  (b)  Simulation results of the fully optimized (zero-temperature) using the optimal algorithm for no bond crossing in Sec. IV. Cost$=-533.967$. %36.033- 570. 
  (c) Distance distributions of the optimized network $P(d^*)$ for the cases in (a) and (b). }
  \label{NMFT=0lam1pov}
\end{figure}

In this section, we examine the properties of the fully optimized network. We shall focus on the situation that the edge weights are the planar distance between two nodes, i.e. $w_\alpha=d_\alpha$. Standard network properties such as clustering coefficients, shortest path lengths, mean degree and degree distribution, and the distribution of the distances of the optimized networks are measured from Monte Carlo simulations, numerical solution of the mean-field equations, and by our optimal algorithm in Sec. IV. $N=100$ nodes are randomly distributed on a unit square, the optimized connectivity configuration is obtained by the methods mentioned above and  various network quantities are measured. We first look into the fully optimized (zero-temperature) network configurations under different crossing penalty, as displayed in Fig. \ref{NMFT=0lam_2pov} for a fixed value of $\lambda$ and the same realization of the nodes. For finite values of $\gamma$, the optimized solution is obtained by solving the mean-field equation numerically and the connections maps are shown in Fig. \ref{NMFT=0lam_2pov}a and \ref{NMFT=0lam_2pov}b. In general long distance connections are less frequent and both  the number of edges and number of crossings  decrease with $\gamma$. Under the strict no bond-crossing constraint, only the optimal algorithm in Sec. IV can find the solution quickly, whose planar network connection map is shown in Fig. \ref{NMFT=0lam_2pov}c. For the case of no bond crossing case, there is almost a full triangulation of the 2D plane, except for a few long distance nodes that are not connected. And if $\lambda$ is sufficiently large, all possible edges with no bond crossing are connected.
For the case of no bond-crossing, one can further compare the resultant planar network configurations obtained by traditional Monte Carlo simulations annealed to a very low temperature and by our optimal algorithm. Fig. \ref{NMFT=0lam1pov}a and \ref{NMFT=0lam1pov}b show the two planar networks of the same node realization obtained from these two methods. Careful examination reveals that the Monte Carlo algorithm gives rise to the unsatisfactory scenario of having some longer distance connections that forbid other shorter distance connections due to the no bond-crossing constraint. The optimized costs are $-527.735$ and $-533.967$ for the networks in Fig. \ref{NMFT=0lam1pov}a and \ref{NMFT=0lam1pov}b respectively, indicating that our optimal algorithm can search a lower cost solution with a very small cost difference (about 1\%). The distance distributions of the optimized network are also shown in Fig. \ref{NMFT=0lam1pov}c, indicating that the optimal algorithm finds  more shorter connections. 

\subsection{Optimized distance and degree distributions}

\begin{figure}[htbp]
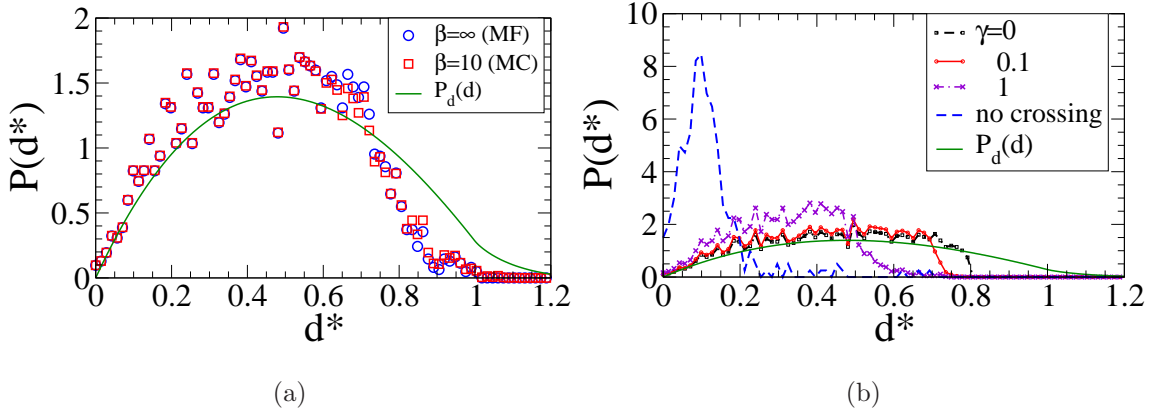

\centering
   \subfigure[]{\includegraphics*[width=\figws]{Pstarlam1gam1.eps}}
\subfigure[]{\includegraphics*[width=\figws]{Pstarlam_4T=0.eps}}
%\subfigure[]{\includegraphics*[width=\figws]{ModCNoXingbeta10lam1Pd.eps}}
  \caption{ Optimized distance distribution function for the optimized network. $N=100$ nodes are distributed uniformly   on a unit square. The distance distribution for all distances between two nodes (Eq. (\ref{Pd}), solid curve) is also shown for comparison.
   (a)  $\lambda=1$, $\gamma=1$ for $\beta=10$ (by Monte Carlo simulations) and zero-temperature (solution of the mean field equation).  (b)  
   The fully optimized (zero-temperature) distance distribution with $\lambda=0.4$ for  different values of $\gamma$. }  \label{Pdstar}
\end{figure}

After the network has been optimized, the distribution of the weighs with $A_{ij}=1$ are then measured to give the optimized degree and weight (distance) distributions. For the case of $c_0=0$ the weights are simply the distances of the connecting nodes. Fig. \ref{Pdstar}a shows the distance distribution, $P(d^*)$, of the optimized network with crossing penalty $\gamma=1$ obtained by Monte Carlo simulations at low temperature ($\beta=10$) and from solution of the zero-temperature mean-field equations. The  original distribution of all the distances between two nodes is also shown for comparison, indicating that the optimized network tends to avoid long distance connections. The effect of crossing penalty on the optimized distance distribution is shown in Fig. \ref{Pdstar}b for the same node realization and $\lambda=0.4$. As $\gamma$ increases, $P(d^*)$ becomes skewed with  substantial decrease    near the tail. And in the no bond-crossing case, $P(d^*)$ becomes a narrow distribution peak at the value of mean nearest inter-node separation, given by $1/\sqrt{N}=0.1$ for $N=100$.
%\begin{figure}[htbp]\begin{center}
%\subfigure[]{\includegraphics*[width=\figwss]{ModCbeta10gam1Pd.eps}}
%  \subfigure[]{\includegraphics*[width=\figwss]{ModCbeta10gam5Pd.eps}} \end{center}
 % \caption{Cumulative optimized weight distribution function for the optimized network at $\beta=10$  with (a) $\gamma=1$, (b) $\gamma=5$.  (c)  }  \label{PcumNoX}\end{figure}

The number of connections or local degree of each node of the optimized network is also measured. The fully optimized degree distributions $P_{opt}(k)$,  for $\lambda=0.4$  with various values of $\gamma$  are shown in Fig. \ref{meank}a. Average over 100 to 200 node realizations are performed to give a smoother $P_{opt}(k)$. 
As $\gamma$ increases, the peak of $P_{opt}(k)$ shifts to lower degrees and for the case of no bond-crossing constraint $P_{opt}(k)$ becomes rather narrow and peak at $k\simeq 6$ (see Fig. \ref{meank}b). The degree distribution for a random network of the same $\bar k$, which is a binomial distribution, is also shown for comparison, indicating that $P_{opt}(k)$ is substantially narrower than that of a random network.

The mean degree $\bar{k}$ of the optimized network can also be measured and some results under no bond-crossing constraints are shown in Fig. \ref{xovsgam}b. Here we  present an analytic result of  $\bar{k}$ for the fully optimized network with no bond-crossing by exploiting results in planar triangulation\cite{Loera2010}, one can show  
 (see Appendix II for a derivation)
\begin{equation}
\bar{k}=6-\frac{16}{3N}\left(\gamma_{E}+\text{ln}\frac{N}{2}\right)-\frac{6}{N}, \label{meandegree}
\end{equation}
where $\gamma_{E}=0.5772156649$ is the Euler's constant.
%Fig. \ref{Figure-15} shows the number of edges ($N_{e}$) and the number triangles ($N_{t}$) of the optimized network with the strictly no bond crossing constraint as a function of the number of nodes ($N$) and the theoretical curves Eqs. (\ref{Net3})(\ref{Net4}).
Fig. \ref{meank}c shows the mean degree $\bar{k}$ vs. $N$ of the optimized network with the strictly no bond crossing constraint measured from simulations using our optimal algorithm together with the theoretical result Eq. (\ref{meandegree}), showing perfect agreement. When the number of nodes $N$=100, $\bar{k}\simeq 5.7$ (which compares well with the result in Fig. \ref{xovsgam}b) and tends to 6 as $N \rightarrow \infty$.
\begin{figure}[htbp]
\centering
\subfigure[]{ \includegraphics[width=\figwss]{PoptkT0lam_4b.eps}}
\subfigure[]{ \includegraphics[width=\figwss]{PkoptT0NBXlam_4.eps}}%{Degreedistribution-Nocrossing-lambda=1}}
\subfigure[]{ \includegraphics[width=\figwss]{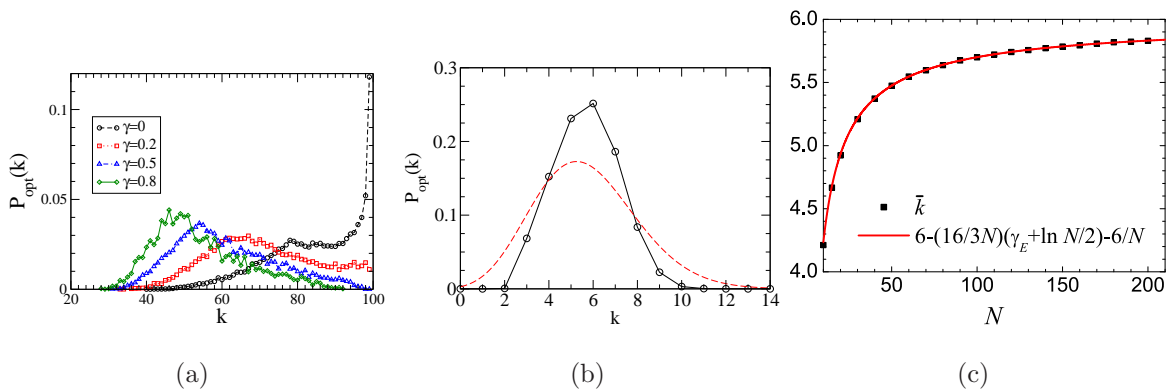}}
\caption{(a)  Degree distributions of the fully optimized network with $\lambda=0.4$ and $\gamma=0, 0.2, 0.5, 0.8$. (b) Degree distributions of the fully optimized network with $\lambda=0.4$ under no bond-crossing constraint. The degree distribution of a random network with the same mean degree ($\bar{k}=5.7$, dashed curve) is also shown for comparison. (c) The mean degree $\bar{k}$ of the optimized network with the strictly no bond crossing constraint vs. $N$. The theoretical  result (\ref{meandegree})  (curve)  is also shown.}
\label{meank}
\end{figure}

\subsection{Clustering Coefficients and Minimal path length}
\begin{figure}[htbp]
\centering
 \subfigure[]{\includegraphics*[width=\figws]{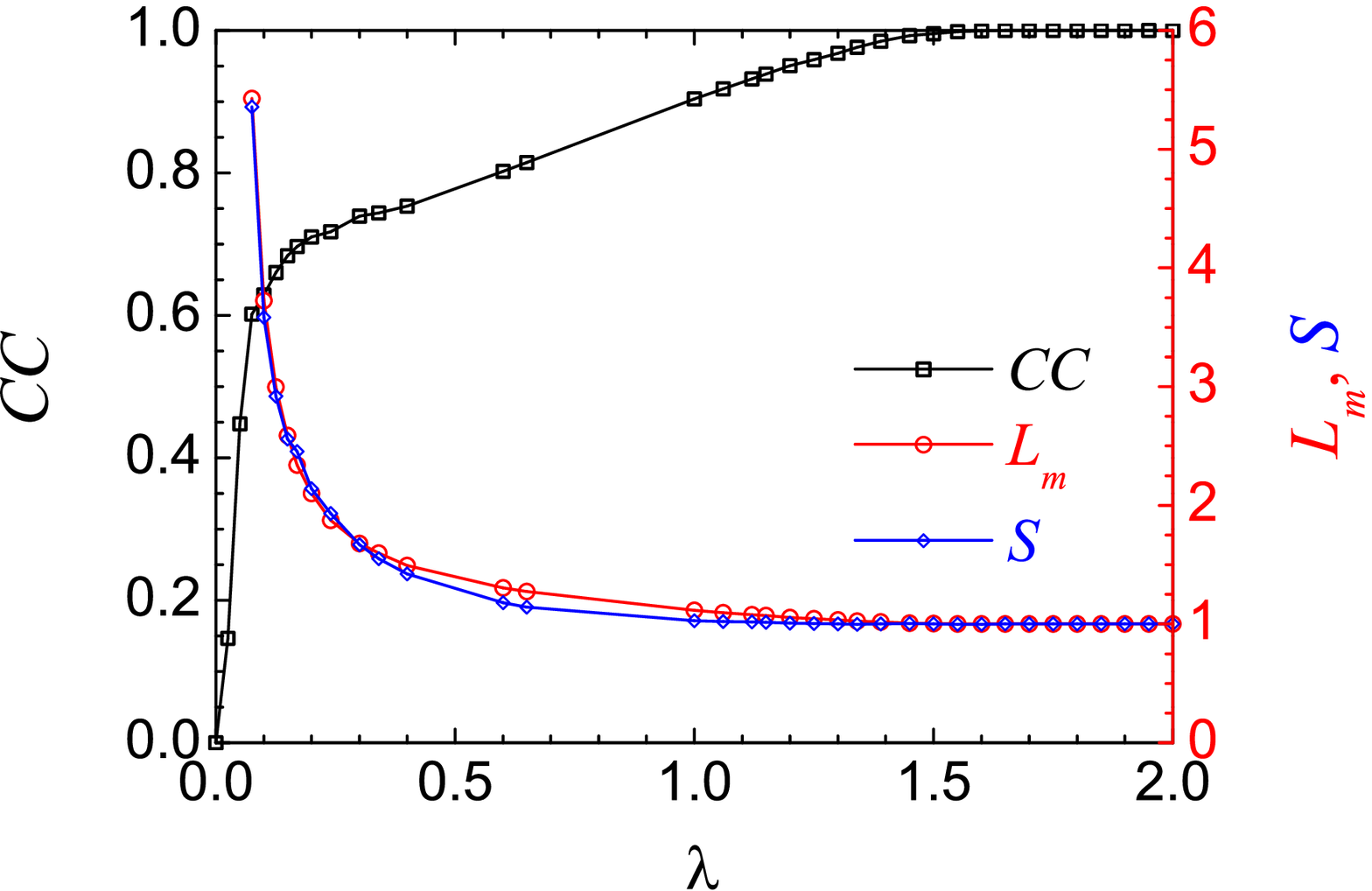}}
 \subfigure[]{\includegraphics*[width=\figws]{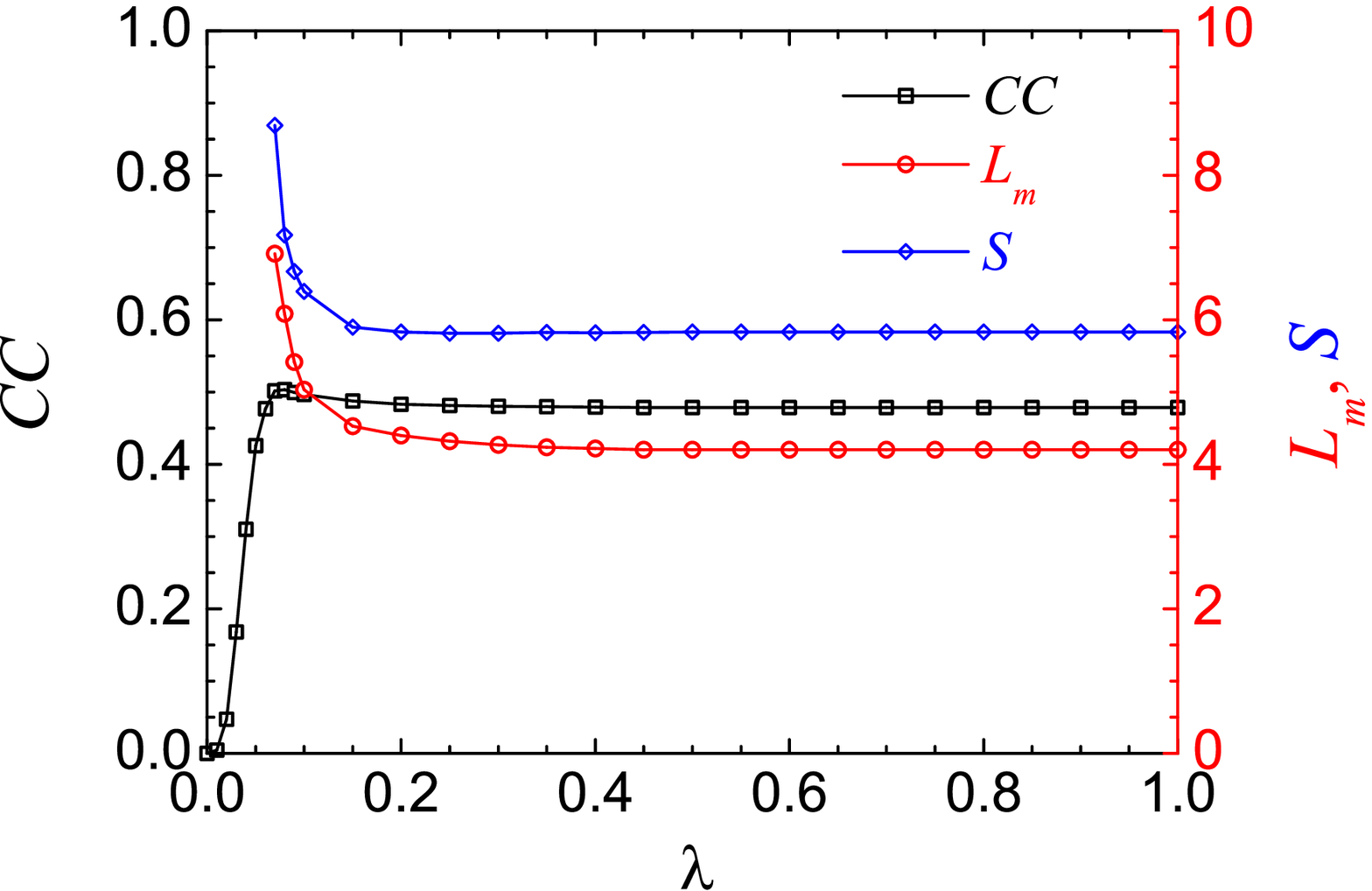}}
% \subfigure[]{\includegraphics*[width=\figws]{ModCSdeggam1_.eps}}
% \subfigure[]{\includegraphics*[width=\figws]{ModCqgam1_.eps}}
\caption{Simulation results  of the  clustering coefficients, average minimal path lengths and small-worldness as a function of $\lambda$ for (a) $\gamma=1$ (by solving the mean-field equations),  and (b) under no bond crossing constraint (using the optimal algorithm for no bond crossing in Sec. IV).} \label{CCLS}\end{figure}

The clustering coefficient ($CC$) and the minimal path length ($L_m$) of the optimized network are also measured from the optimized network configurations.  Fig. \ref{CCLS} plots the $CC$ and $L_m$ vs. $\lambda$ for  $\gamma=1$ and under the no bond-crossing constraint. For finite $\gamma$, $CC$ increases with $\lambda$ and saturates to unity for sufficiently large $\lambda$ as the network tends to be completely connected, at the same time $L_m$ decreases and approaches unity. On the other hand, for the case of no bond-crossing, $CC$ increases and rapidly saturates to a value of $\sim 0.5$ and $L_m$ decreases to about 4, for moderate values of $\lambda$. The relatively low value of $CC$ can be attributed to the fact that the optimized network  is  filled with planar triangulation for sufficiently large $\lambda$.       A network with a large $CC/L_m$ ratio would possesses small-world characteristics\cite{strogatz98} which can be quantified by the small-worldness, $S$,  defined as the ratio of $CC/L_m$ relative to that of a  random network\cite{ER,random} of the same mean degree,
\begin{align}
S=\frac{\frac{CC}{CC_{rand}}}{\frac{L_{m}}{L_{m_{rand}}}},
\end{align}
where $C_{rand}$ and $L_{m_{rand}}$ are the clustering coefficient and the average minimal path length of the random network with the same mean degree. 
The small-worldness of the optimized network is also plotted in Fig. \ref{CCLS}, indicating  a small-world signature of the optimized network with $S\simeq 6$ under the no bond-crossing constraint.

\section{Summary and Outlook}
 In this paper,  we investigated in detail on the problem of network connection growth for nodes embedded on a two-dimensional plane that aims to maximize connections between the nodes but also minimize connection cost with edge crossing penalized. By mapping the network links to Ising spins, the system can be investigated in the  statistical mechanics framework of  a dilute anti-ferromagnetic spin system whose  low-temperature (fully optimized) behavior is dominated by  frustrations which originated from the edge crossing penalty.  Using  mean field theories, we derived analytic results  for the order-parameter (related to the mean degree of the network) and the associated phase diagram under the crude mean-field approximation. The results are also verified  explicitly by  Monte Carlo simulations. The crude mean-field approximation give rather satisfactory results except for strong bond crossing penalty cases. The optimized network configuration can be obtained from the numerical solutions of the full local mean-field equations with the optimized solution chosen to be the lowest cost solution found. However, as bond-crossing penalty that causes the frustration of the system to be stronger, finding the solutions of the mean-field equations becomes a formidable task in practice, even for systems of moderate sizes, as the cost landscape gets more complex and rugged. The complex cost landscape makes the traditional Monte Carlo algorithm to search for the optimized (ground) state difficult as in other frustrated spin systems\cite{SG,graph87,graph88}. Hence, for the challenging case of no bond-crossing characterized by strong frustrations, based on the mean-field equations, we further developed an efficient algorithm to find the optimal network configurations that can satisfactorily find the optimized network configurations.
It is worth to note that in the limit of strictly no bond-crossing, our model in the large $\lambda$ limit can be  mapped exactly to the problem of minimal length planar triangulation.
 Finding the optimized network with the strictly no bond crossing constraint and maximal possible connections is equivalent to the problem of minimum length triangulation, which has been shown to be a NP-hard problem\cite{Loera2010,Rote2008}.
 
There are some possible improvement or extension for the present approach. 
For example, the crude mean-field approximation might be improved by using the TAP equation\cite{TAP} as in quenched disordered systems such as spin glasses. One can also consider optimization by rewiring the edges to conserved the total number of connections, this can be modeled by the spin exchange dynamics (Kawasaki dynamics) with conserved order parameter\cite{kawasaki}.
Also one can  extend the model to take into account for the situations that the connection formation tendency is weighted with a weight  that are different for each nodes\cite{cheng}. This would be relevant in designing roadway systems in which a city or township (nodes) with a higher population will be given a higher node weight for making connections. Furthermore, this work considered only undirected connection between nodes, it may be possible to extend the framework to consider connections to be directed  for optimized directed networks which is of particular relevance in construction of directed roadways  or for circuits with rectified directional currents.
 
One ambitious  goal is to deduce the evolution rule, presumably governed by some optimization model, from the observed network structure. This is a challenging inverse problem of inferring a cost function and its parameters from the observed connections (such as $w_\alpha^{opt}$). This problem is even more difficult in practice since often the network connection weights and directions are not directly observable or known, and  in many cases only  the dynamical data of network nodes can be observed. One needs to first infer the network structure from the time-series data of the nodes. With the dynamical data of network nodes are increasingly available, there are reliable methods  for the reconstruction of  networks from the time-series dynamics data of the nodes\cite{CLL2013,CLL2015,zhang2015,CT2017,Lai2017,TCL,Lai2019,Lai2021} that can be employed. Hopefully with suitable implementation of the network reconstruction schemes from dynamical data and the inference of optimization model from network structure to be developed, then it would open a new avenue of deducing the network evolution rule (which is usually a long time scale process) from the observed node dynamical data in a relatively short observation period.
 
\subsection*{Appendix I: Derivation of $P_d$ for uniformly  distributed nodes in a unit square} %Pd.f
\begin{figure}[htbp]
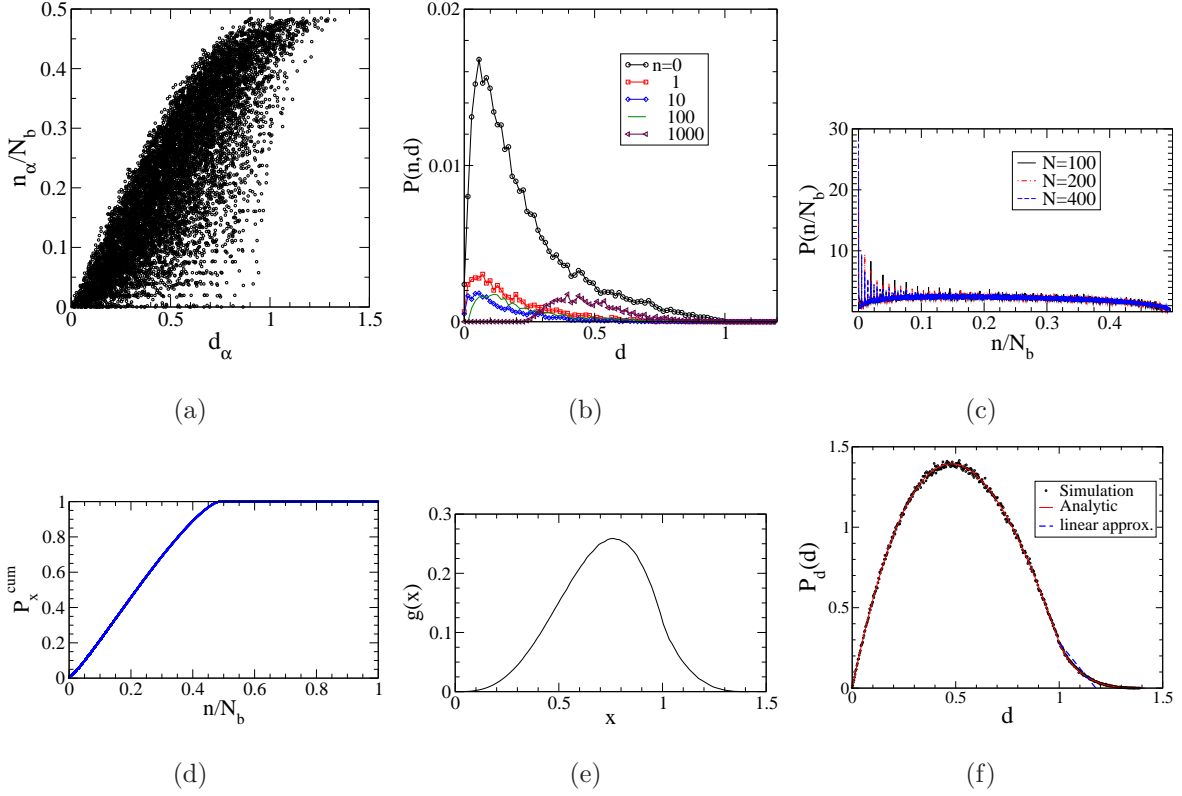

\centering
 \subfigure[]{  {\includegraphics*[width=\figwss]{nvsdN100.eps}}} 
  \subfigure[]{  {\includegraphics*[width=\figwss]{PndN=100.eps}}}   
    \subfigure[]{  {\includegraphics*[width=\figwss]{Pn_Nb.eps}}}
%  \subfigure[]{  {\includegraphics*[width=\figwss]{Pn_Nbzoom.eps}}}
 \subfigure[]{  {\includegraphics*[width=\figwss]{PxcumvsnN100.eps}}} 
 \subfigure[]{  {\includegraphics*[width=\figwss]{crudeMFgx.eps}}}
  \subfigure[]{  {\includegraphics*[width=\figwss]{Pd2.eps}}}      
  \caption{ Various distributions about the distances between two points and number of crossings between possible dges for $N$ point uniform randomly distributed  on a unit square. (a) The (nomralized) number of possible crossings of an edge plotted against its length for  $N=100$ points. (b) The probability of an edge of length $d$ and having $n$ crossings $P(n,d)$ for  $N=100$ points, plotted as a function of $d$ for $n=0$, 1, 10, 100 and 1000.     (c) Probability distribution  of the number of crossings ($n$) normalized by the maximal possible crossings ($N_b$). Each distribution is obtained with an average of 100 realizations. $N=100$, 200 and 400. (d) 
 % Magnification of (c) for $n$ near zero showing  finite number of edges with no crossing persists as $N$ increases. (e)
  The cumulative distribution of the number of crossings. (e) The function $g(x)\equiv \dfrac{x}{N_b} \sum_{n=0}^{N_b} n P(n,x)$ for  $N=100$ points on a unit square. (f) Distribution of the distance between two points uniformly and randomly distributed in a unit square. Results obtained by simulations and analytic formula (\ref{Pd}) are both displayed. The dashed straight line shows the linear approximation of the tail.}\label{Pdfig}
  \end{figure}

Here we derive the analytical expression for the distribution of the distances between two points uniformly and randomly  distribution in a unit square. Suppose two points with coordinates $(x_1,y_1)$ and $(x_2,y_2)$ are randomly chosen in $[0,1)\times[0,1) $, we first compute the distribution of the separation between $x_1$ and $x_2$ to give
\begin{eqnarray}
& &\int_0^1\int_0^1 \delta(u-|x_1-x_2|)dx_1 dx_2= 2(1-u), \quad \mbox{for $u\in [0,1)$ and 0 otherwise.}
\end{eqnarray}
The same result for the separation between $y_1$ and $y_2$ can be obtained. Then the distribution for the square of the distance is given by
\begin{eqnarray}
P_{d^2}(w)&=&4\int_0^1\int_0^1 (1-u)(1-v)\delta(w-u^2-v^2)\\
&=&\left\{
\begin{array}{ll}
\pi-4\sqrt{w}+w     , & \mbox{$0<w\leqslant 1$}\\
 2[\csc^{-1}\sqrt{w}-\tan^{-1}\sqrt{w-1}-1] +4\sqrt{w-1}-w ,& \mbox{$1<w<{2} $}.
\end{array}\right.
\end{eqnarray}
Finally the distribution of the distance, $P_d$, can be obtained upon changing the variable $X=\sqrt{w}$ to give (\ref{Pd}).
Fig. \ref{Pdfig}f displays $P_d$ measured from Monte-Carlo simulations together with the analytic expression (\ref{Pd}) showing perfect agreement.
 The functional form of $P_d(X)$ for $X>1$ is complicated for further analytic calculations. One can approximate it by a linear form by requiring that the total probability (area under the curve) for $1<X\leqslant d_m$ to be the same as the original one ($={19 \over 6}-\pi$), where $d_m$ is the maximum value of the distance in this approximation. Such a linear approximation for the tail of $P_d$ is
  \begin{equation}
 P_d(X)\simeq P_d(1)\left(\frac{d_m-X}{d_m-1}\right)\qquad 1<X\leqslant d_m,
 \end{equation}
  where $P_d(1)=2(\pi-3)$ and $d_m=\frac{1}{6(\pi-3)}$, which is also displayed in Fig. \ref{Pdfig}f (dashed line) for comparison.
  
\subsection*{Appendix II: Relation to Triangulation of a planar graph}
In graph theory, the problem of triangulation is to connect the given nodes on a plane resulting in a partition of the plane into triangles. The triangulation of a point set of $N$ points in a two dimension plane has the following exact results for the number of edges and number of triangles\cite{Loera2010}
\begin{eqnarray}
&&N_{e}=3N-h-3  \quad \text{edges}  \label{Net1}\\
&&N_t=2N-h-2  \quad \text{triangles}, \label{Net2}
\end{eqnarray}
where $h$ is the number of points in the boundary of the convex hull of the point set. For a given location of  the points, one can count the number of points in the convex hull ($h$) so the number of edges ($N_{e}$) and number of triangles ($N_{t}$) of the triangulation can be determined.
For $w_\alpha=d_\alpha$, the cost in (\ref{HC}) can be written in terms of the number of edges and number of crossings
\begin{equation}
{\cal C_X} = \sum_{\alpha}A_{\alpha}d_{\alpha} -2\lambda N_e+\frac{4 \gamma}{N_b} N_x .\label{CX}
\end{equation}
Hence finding the optimized network with the strictly no bond crossing ($N_x=0$) constraint for sufficiently large $\lambda$, is equivalent to problem of finding the minimum length triangulation. It has been shown that finding the minimum length triangulation is a NP-hard problem\cite{Loera2010,Rote2008}. In addition, the number of possible triangulations is equal to the number of solutions of the mean-field equations in which the parameters satisfy the condition given by (\ref{Nocrossing}). It is also known that the number of triangulations grows exponentially with $N$\cite{Loera2010}, and hence so does the number of mean-field solutions (in the no bond-crossing limit).

One can estimate $N_{e}$ and $N_{t}$ of the optimized network with the strictly no bond crossing constraint as a function
of the $N$ using Eqs. (\ref{Net1})(\ref{Net2}). If the nodes set are uniformly random distributed in a square, the expected number of nodes that lie in the convex hull is given by\cite{Weisstein, Finch2003}
\begin{equation}
\lim_{N \to \infty}  h\simeq\frac{8}{3}(\gamma_{E}-\text{ln}2+\text{ln}N), \label{Nch}
\end{equation}
where $\gamma_{E}\simeq$=0.577216 is the Euler–Mascheroni constant. From Eqs.  (\ref{Net1}),  (\ref{Net2}) and(\ref{Nch}), $N_{e}$ and $N_{t}$ can be approximated as
\begin{eqnarray}
&&N_{e}\simeq 3N-\frac{8}{3}\left(\gamma_{E}+\text{ln}\frac{N}{2}\right)-3 \label{Net3}\\
&&N_{t}\simeq 2N-\frac{8}{3}\left(\gamma_{E}+\text{ln}\frac{N}{2}\right)-2 . \label{Net4}
\end{eqnarray}
Hence one obtain an analytic expression for the mean degree $\bar{k}$ for triangulation (network with maximal connections with no edge-crossing)
\begin{equation}
\bar{k}=\frac{2N_{e}}{N}=6-\frac{16}{3N}\left(\gamma_{E}+\text{ln}\frac{N}{2}\right)-\frac{6}{N}. \label{meandegree2}
\end{equation}

\begin{acknowledgments}  This work  has been supported by Ministry of Science and Technology of Taiwan
 under the grant no.  MOST 110-2112-M-008-026-MY3, and NCTS of Taiwan.
\end{acknowledgments}

\end{document}